\RequirePackage{lineno}
\documentclass[prd,preprint,tightenlines,superscriptaddress,showpacs]{revtex4}
\usepackage{graphicx}
\usepackage{epsfig}
\usepackage{dcolumn}
\usepackage{bm,color}
\usepackage{multirow}

\newcommand{\eff}{\varepsilon}
\newcommand{\BR}{{\cal B}}

\newcommand{\piz}{\pi^0}

\newcommand{\etac}{\eta_c}
\newcommand{\hc}{h_c}

\newcommand{\psp}{\psi(3686)}

\newcommand{\jpsi}{J/\psi}
\newcommand{\ks}{K_{S}^{0}}
\newcommand{\EE}{e^+e^-}

\newcommand{\pp}{\pi^+\pi^-}
\newcommand{\kk}{K^+K^-}

\newcommand{\kskp}{K^{0}_S K^{\pm}\pi^{\mp}}
\newcommand{\kkp}{K^{+} K^{-}\pi^{0}}
\newcommand{\kskppp}{K^{0}_{S} K^{\pm}\pi^{\mp}\pi^{+}\pi^{-}}
\newcommand{\kkppp}{K^{+} K^{-} \pi^{+} \pi^{-} \pi^{0}}
\newcommand{\kktp}{K \bar{K} \pi \pi \pi}

\newcommand{\ccj}{\chi_{cJ}}
\newcommand{\ccz}{\chi_{c0}}
\newcommand{\cco}{\chi_{c1}}
\newcommand{\cct}{\chi_{c2}}
\newcommand{\beq}{\begin{equation}}
\newcommand{\eeq}{\end{equation}}
\newcommand{\bitm}{\begin{itemize}}
\newcommand{\eitm}{\end{itemize}}
\newcommand{\mev}{\mathrm{MeV}}

\newcommand{\mevcc}{\mathrm{MeV}/c^2}
\newcommand{\gev}{\mathrm{GeV}}
\newcommand{\gevc}{\mathrm{GeV}/c}
\newcommand{\gevcc}{\mathrm{GeV}/c^2}

\begin{document}
\linenumbers
\preprint{} \preprint{\vbox{ \hbox{   }
        \hbox{Intended for {\it Phys. Rev. D}}
        \hbox{Authors: Y. P. Guo, C. Z. Yuan, C. X. Yu}
        \hbox{Committee: M. Maggiora (Chair), F. Liu, X. T. Huang} }}

\title{\quad\\[1.0cm]\boldmath Search for hadronic transition $\ccj\to \etac\pp$ and observation
of $\ccj\to K\bar{K}\pi\pi\pi$}

\author{\small
M.~Ablikim$^{1}$, M.~N.~Achasov$^{5}$, D.~J.~Ambrose$^{39}$, F.~F.~An$^{1}$, Q.~An$^{40}$, Z.~H.~An$^{1}$, J.~Z.~Bai$^{1}$, Y.~Ban$^{27}$, J.~Becker$^{2}$, M.~Bertani$^{18A}$
% ...
\\
\vspace{0.2cm}
(BESIII Collaboration)\\
\vspace{0.2cm} {\it
$^{1}$ Institute of High Energy Physics, Beijing 100049, P. R. China\\
$^{2}$ Bochum Ruhr-University, 44780 Bochum, Germany\\
% ...
$^{5}$ G.I. Budker Institute of Nuclear Physics SB RAS (BINP), Novosibirsk 630090, Russia\\
% ...
$^{18}$ (A)INFN Laboratori Nazionali di Frascati, Frascati, Italy; (B)INFN and University of Perugia, I-06100, Perugia, Italy\\
% ...
$^{27}$ Peking University, Beijing 100871, P. R. China\\
% ...
$^{39}$ University of Rochester, Rochester, New York 14627, USA\\
$^{40}$ University of Science and Technology of China, Hefei 230026, P. R. China\\
% ...
}
\vspace{0.4cm}}



\begin{abstract}

Hadronic transitions of $\ccj\to \etac\pp$~($J=0$, 1, 2) are searched
for using a sample of $1.06\times 10^{8}$ $\psp$ events collected with
the BESIII detector at the BEPCII storage ring. The $\etac$ is
reconstructed with $\kskp$ and $\kkp$ final states. No signals are
observed in any of the three $\ccj$ states in either $\etac$ decay
mode. At the 90\% confidence level, the upper limits are determined to
be $\BR(\ccz\to \etac\pp)<0.07\%$, $\BR(\cco\to \etac\pp)<0.32\%$, and
$\BR(\cct\to
\etac\pp)<0.54\%$. The upper limit of $\BR(\cco\to \etac\pp)$ is
lower than the existing theoretical prediction by almost an order of
magnitude. The branching fractions of $\ccj\to \kskppp$, $\kkppp$,
$\omega\kk$ and $\phi\pp\piz$~($J=0$, 1, 2) are measured for the first
time.

\end{abstract}


\maketitle

\section{Introduction}

Heavy quarkonia, both $c\bar{c}$ and $b\bar{b}$ bound states, have
provided good laboratories for the study of the strong
interaction~\cite{intro-theory2,heavy-quarkonium}. For the hadronic
transitions between the heavy quarkonium states, Yan~\cite{intro-yan}
characterized it as the emission of two soft gluons from the heavy
quarks and the conversion of gluons into light hadrons. Based on this
scheme, a series of decay rates, such as the E1-E1 hadronic
transition, E1-M1 hadronic transition, M1-M1 hadronic transition have
been calculated~\cite{intro-KY,theoreticalchib}. It has been shown
that the multipole expansion can make quite successful predictions for
many hadronic transitions between the heavy
quarkonia~\cite{intro-theory2,intro-theory1}. However, most of these
studies are for the transitions among the ${}^{3}S_{1}$ states; the
hadronic transitions of ${}^{3}P_{J}$ states are seldom
explored. Using a sample of $\Upsilon(3S)$, CLEO measured for the
first time the transition rate of $P$-wave bottomonium
$\chi_{bJ}(2P)\to \chi_{bJ}(1P)\pi\pi$~\cite{CLEOchib}, and the
results are consistent with the theoretical
predictions~\cite{theoreticalchib}. For the hadronic transition of the
$P$-wave charmonium states, there is only an upper limit of 2.2\% at
the 90\% confidence level (C.L.) on the $\cct\to \etac\pp$ transition
rate recently reported by the BaBar experiment~\cite{chic2Babar}. The
most promising process $\chi_{c1}\to \etac\pi\pi$, which is dominated
by an E1-M1 transition, is calculated in the multipole expansion
formalism, and a transition rate $\BR(\chi_{c1}\to \etac\pi\pi) =
(2.72\pm 0.39)$\% is predicted~\cite{theoreticalchic}.

In this article, we search for $\ccj\to \etac\pp$ with $\etac$
decays into $\kskp$ and $\kkp$, where the $\ks$ is reconstructed
in $\pp$ and $\piz$ in $\gamma\gamma$ final states. We also report
the first measurement of the branching fractions of $\ccj\to
\kskppp$, $\kkppp$, $\omega\kk$, and $\phi\pp\piz$.

\section{The experiment and data sets}

The data sample for this analysis consists of $1.06\times 10^{8}$
events produced at the peak of the $\psp$ resonance~\cite{npsp};
an additional 42~pb$^{-1}$ of data were collected at a
center-of-mass energy of $\sqrt{s}$=3.65~GeV to determine
non-resonant continuum background contributions. The data are
accumulated with the BESIII detector operated at the BEPCII $\EE$
collider.

The BESIII detector, described in detail in Ref.~\cite{bes3}, has an
effective geometrical acceptance of 93\% of 4$\pi$. It contains a
small cell helium-based main drift chamber (MDC) which provides
momentum measurements of charged particles; a time-of-flight system
(TOF) based on plastic scintillator which helps to identify charged
particles; an electromagnetic calorimeter (EMC) made of CsI (Tl)
crystals which is used to measure the energies of photons and provide
trigger signals; and a muon system (MUC) made of Resistive Plate
Chambers (RPC). The resolution of the charged particles is $0.5$\% at
$1~\gevc$ in a 1~Tesla magnetic field. The energy loss
($d\mathrm{E}/dx$) measurement provided by the MDC has a resolution
better than 6\% for electrons from Bhabha scattering. The photon
energy resolution can reach $2.5$\% ($5$\%) at $1~\gev$ in the barrel
(endcaps) of the EMC. And the time resolution of TOF is $80$~ps in the
barrel and $110$~ps in the endcaps.

Monte Carlo~(MC) simulated events are used to determine the
detection efficiency, optimize the selection criteria, and study
the possible backgrounds. The simulation of the BESIII detector is
{\sc geant4}~\cite{geant4} based, where the interactions of the
particles with the detector material are simulated. The $\psp$
resonance is produced with \textsc{kkmc}~\cite{KKMC}, while the
subsequent decays are generated with {\sc EvtGen}~\cite{EvtGen}.
The study of the background is based on a sample of $10^8$ $\psp$
inclusive decays which are generated with known branching
fractions taken from the Particle Data Group (PDG)~\cite{pdg}, or
with {\sc lundcharm}~\cite{Lundcharm} for the unmeasured decays.

\section{Event selection}

A charged track should have good quality in the track fitting and be
within the angle coverage of the MDC, $|\cos\theta|<0.93$. A good
charged track (excludes those from $\ks$ decays) is required to be
within 1~cm of the $\EE$ annihilation interaction point (IP)
transverse to the beam line and within 10~cm of the IP along the beam
axis. Charged-particle identification (PID) is based on combining the
$d\mathrm{E}/dx$ and TOF information in the variable $\chi^{2}_{{\rm
PID}}(i)= (\frac{d\mathrm{E}/dx_{\rm measured}-d\mathrm{E}/dx_{\rm
expected}} {\sigma_{d\mathrm{E}/dx}})^2+ (\frac{{\rm TOF}_{\rm
measured}-{\rm TOF}_{\rm expected}}{\sigma_{\rm TOF}})^2$. The values
$\chi^{2}_{{\rm PID}}(i)$ and the corresponding confidence levels
${\rm Prob_{PID}}(i)$ are calculated for each charged track for each
particle hypothesis $i$ (pion, kaon, or proton).

Photons are reconstructed from isolated showers in the EMC which
are at least 20 degrees away from any of the charged tracks. In
order to improve the reconstruction efficiency and the energy
resolution, the energy deposited in the nearby TOF counter is
included. Photon candidates are required to have the energy
greater than 25~$\mev$ in the EMC barrel region
($|\cos\theta|<0.8$), while in the EMC endcap region
($0.86<|\cos\theta|<0.92$), the energy threshold requirement is
increased to 50~$\mev$. EMC timing requirements are used to
suppress noise and energy deposits unrelated to the event.

$\ks$ candidates are reconstructed from secondary vertex fits to all
the charged-track pairs in an event (assume the tracks to be $\pi$). The
combination with the best fit quality is kept, and the $\ks$ candidate
must have an invariant mass within $7~\mevcc$ of the $\ks$ nominal
mass and the secondary vertex be at least 0.5~cm away from the IP. The
reconstructed $\ks$ information is used as input for the subsequent
kinematic fit. The $\piz$ candidates are reconstructed from pairs of
photons with an invariant mass in the range $0.120~\gevcc
<M(\gamma\gamma)< 0.145~\gevcc$.

In selecting $\psp\to \gamma\kskppp$, a candidate event should
have at least six charged tracks and at least one good photon.
After $\ks$ selection, the event should have exactly four
additional good charged tracks with zero net charge. While in
selecting $\psp\to \gamma\kkppp$, a candidate event should
have four good charged tracks with zero net charge and at least
three good photons. The $\gamma\kskppp$ ($\gamma\kkppp$) candidate
is then subjected to a four-constraint (4C) kinematic fit to reduce
background and improve the mass resolution. Determination of the
species of the final state particles and selection of the best
photons when additional photons (and $\piz$ candidates) are found
in an event are achieved by selecting the combination with the
minimum value of $\chi^{2}=\chi^{2}_{\rm 4C} +
\sum_{j=1}^{4}\chi^{2}_{\rm PID}(j)$, where $\chi^{2}_{\rm 4C}$ is
the chi-square from the 4C kinematic fit. Events with $\chi^{2}_{\rm
4C}<50$ are kept as $\gamma\kskppp$ ($\gamma\kkppp$) candidates.

There is substantial background from $\psp\to X+\jpsi$ decays.  The
$\psp\to \pp\jpsi$ events are removed by requiring the recoil mass of
any $\pp$ pair to be outside of a $\pm 3\sigma$ window around the
$\jpsi$ nominal mass, where $\sigma$ is the resolution of the $\pp$
recoil mass. The $\psp\to\eta\jpsi$, $\eta\to\gamma\pp$ events are
rejected if $0.535~\gevcc<M(\gamma\pp)<0.555~\gevcc$.  To reject
background with one more photon than the signal events
($\gamma\gamma\kskppp$ or $\gamma\gamma\kkppp$), $\chi^{2}_{\rm
4C}>10$ is required when a 4C kinematic fit to $\gamma\gamma\kskppp$
or $\gamma\gamma\kkppp$ is applied to the event; and to suppress
$\psp\to \piz\piz\jpsi,~\jpsi\to\kk\pp$ background,
$M(\kk\pp)>3.125~\gevcc$ or $M(\kk\pp)<3.095~\gevcc$ is required.

\section{Data analysis}

\subsection{\boldmath $\ccj\to\kskppp$ and $\ccj\to\kkppp$}

After the above selection, the invariant mass distributions
of the hadron system are shown in Fig.~\ref{fig:invariant mass
kkpipipi}, and clear $\ccj$ signals are observed with very low
background level in the two modes.

\begin{figure*}[htbp]
\begin{center}
\includegraphics[width=3.0in,height=2.0in]
{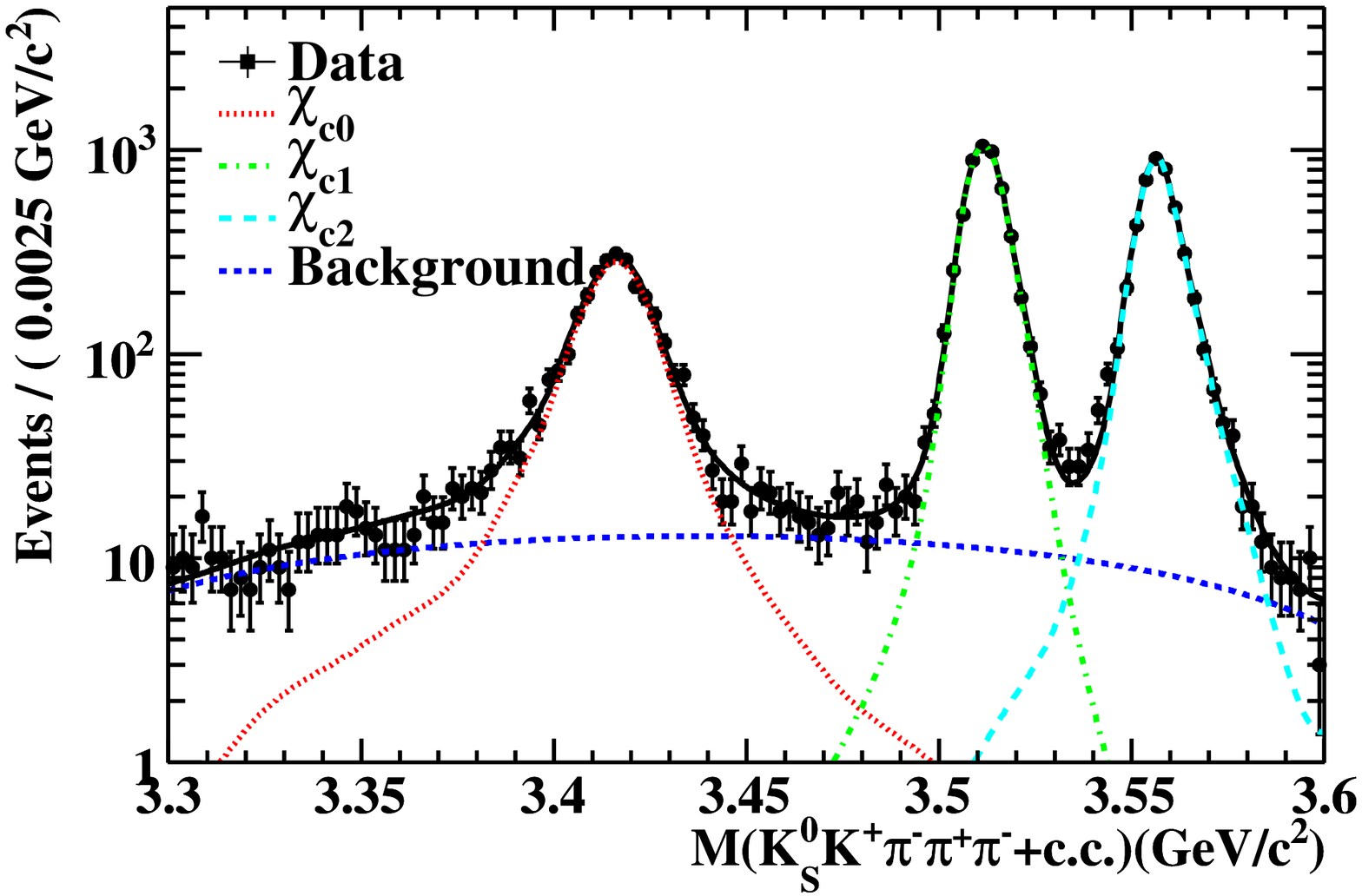}
\includegraphics[width=3.0in,height=2.0in]
{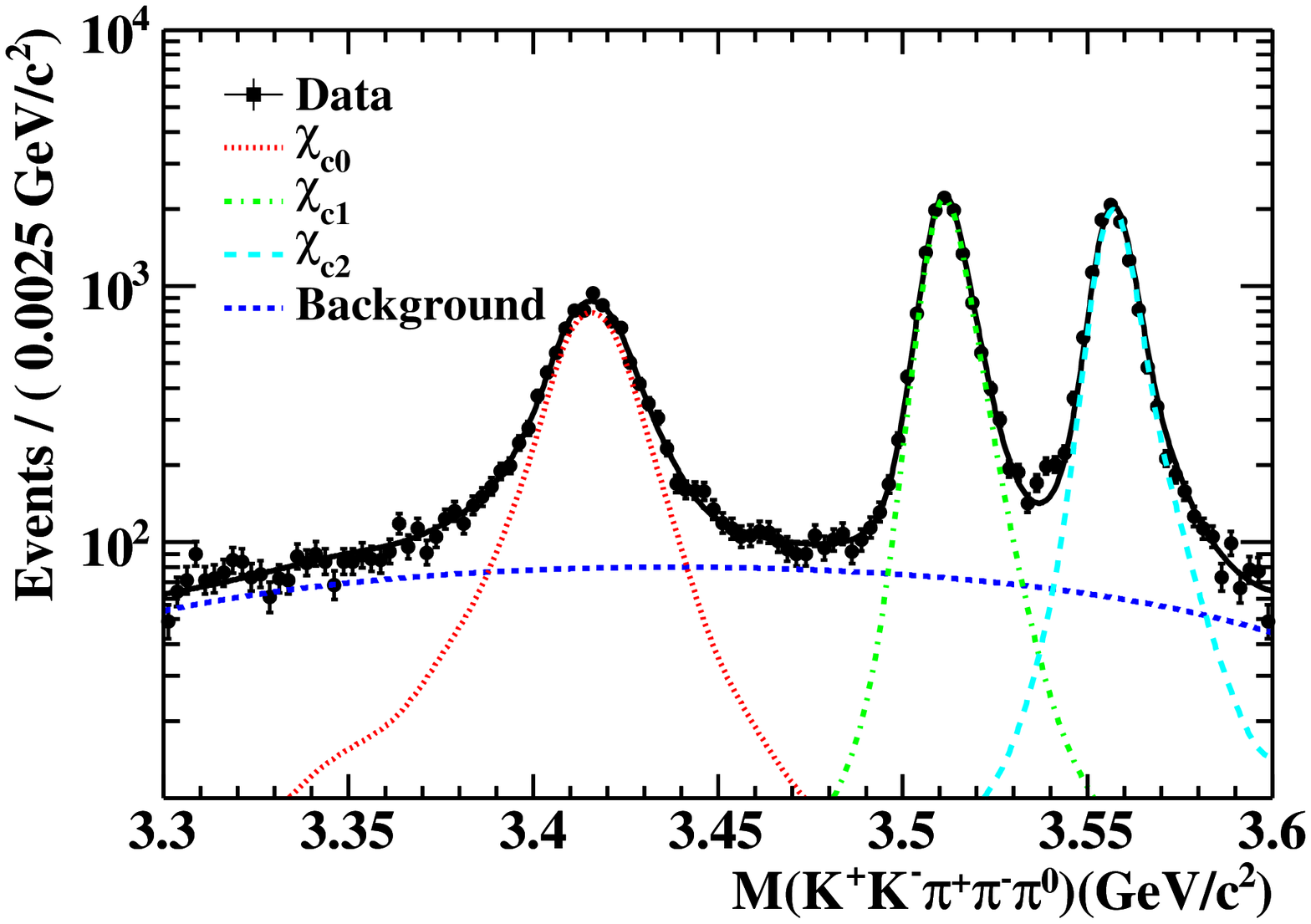}
\end{center}
\caption{Invariant mass spectrum of $\kskppp$ (left panel) and
$\kkppp$ (right panel), together with the best fit results. The
points with error bars are data, and the solid lines are the total fit
results. The $\ccz$, $\cco$, and $\cct$ signals are shown as
dotted lines, dash-dotted lines, and long-dashed lines,
respectively; the backgrounds are in dashed lines.}
\label{fig:invariant mass kkpipipi}
\end{figure*}

Using the inclusive MC events sample, the potential backgrounds from
the $\psp$ decays which may contaminate $\ccj\to\kskppp$
($\ccj\to\kkppp$) are estimated. Events from $\psp\to \gamma\ccj$,
$\ccj\to \ks\ks\pp$ and $\ccz\to \ks\ks\kk$ ($\psp\to \gamma\ccj$,
$\ccj\to \ks K^{\pm}\pi^{\mp}\piz$) may peak at the signal region, and
the contributions from these peaking backgrounds are estimated using
the detection efficiencies determined from the MC simulations and the
corresponding branching fractions from previous
measurements~\cite{pdg}, and then subtracted. These background events
after final event selection are listed in Table~\ref{table:peaking
bkg ksk3pi} and ~\ref{table:peaking bkg 2k2pipi0}; the errors in the numbers of events are from the uncertainties
of the detection efficiency and branching fractions. The other
backgrounds are composed of dozens of decay modes and smoothly
distributed in the full mass region ($3.3~\gevcc\sim 3.6~\gevcc$);
this kind of background is described by a second-order Chebyshev
function.

\begin{table*}[htbp]
\caption{The number of remanent peaking background events ($N_{\rm
bkg}^{\rm peak}$) in $\ccj\to\kskppp$ after final
event selection. The branching fractions ($\mathcal{B}$) are taken
from PDG~\cite{pdg}.} \label{table:peaking bkg ksk3pi}
\begin{center}
\begin{tabular}{ccc}
\hline Decay modes   & ~~~~~~~~~~$N_{\rm bkg}^{\rm peak}$~~~~~~~~
&~~~~$\mathcal{B}$~\cite{pdg}~~~  \\\hline
$\ccz\to\ks\ks\pp$       &  $4.9\pm 1.4$   &  $(5.8\pm1.1)\times10^{-3}$\\
$\cco\to\ks\ks\pp$       &  $0.1\pm 0.1$   &  $(7.2\pm3.1)\times10^{-4}$ \\
$\cct\to\ks\ks\pp$       &  $0.6\pm 0.3$   &  $(2.4\pm0.6)\times10^{-3}$  \\
\hline
$\ccz\to\ks\ks\kk$       &  $43.3\pm 15.8$ & $(1.4\pm0.5)\times10^{-3}$  \\
\hline
\end{tabular}
\end{center}
\end{table*}

\begin{table*}[htbp]
\caption{The number of remanent peaking background events ($N_{\rm
bkg}^{\rm peak}$) in $\ccj\to\kkppp$ after final
event selection. The branching fractions ($\mathcal{B}$) are taken
from PDG~\cite{pdg}.} \label{table:peaking bkg 2k2pipi0}
\begin{center}
\begin{tabular}{ccc}
\hline Decay modes   & ~~~~~~~~~~$N_{\rm bkg}^{\rm peak}$~~~~~~~~
&~~~~$\mathcal{B}$~\cite{pdg}~~~  \\\hline
$\ccz\to\ks K^{\pm}\pi^{\mp}\piz$       &  $123.8\pm 19.4$ &  $\frac{1}{2}\times(2.52\pm0.34)\times10^{-2}$\\
$\cco\to\ks K^{\pm}\pi^{\mp}\piz$       &  $39.0\pm9.1$  &  $\frac{1}{2}\times(9.0\pm1.5)\times10^{-3}$  \\
$\cct\to\ks K^{\pm}\pi^{\mp}\piz$       &  $52.5\pm 9.3$  &  $\frac{1}{2}\times(1.51\pm0.22)\times10^{-2}$  \\
\hline
\end{tabular}
\end{center}
\end{table*}

Data taken at $\sqrt{s}=3.65~\gev$ are used to estimate
backgrounds from the continuum process $\EE\to q\bar{q}$. This kind of
background is found to be small and uniformly distributed in the
full mass region of interest in both decay modes, so the
contribution can be represented by the smooth background term in
the fit.

\begin{figure*}[htbp]
\includegraphics[width=2.5in,height=2.0in]{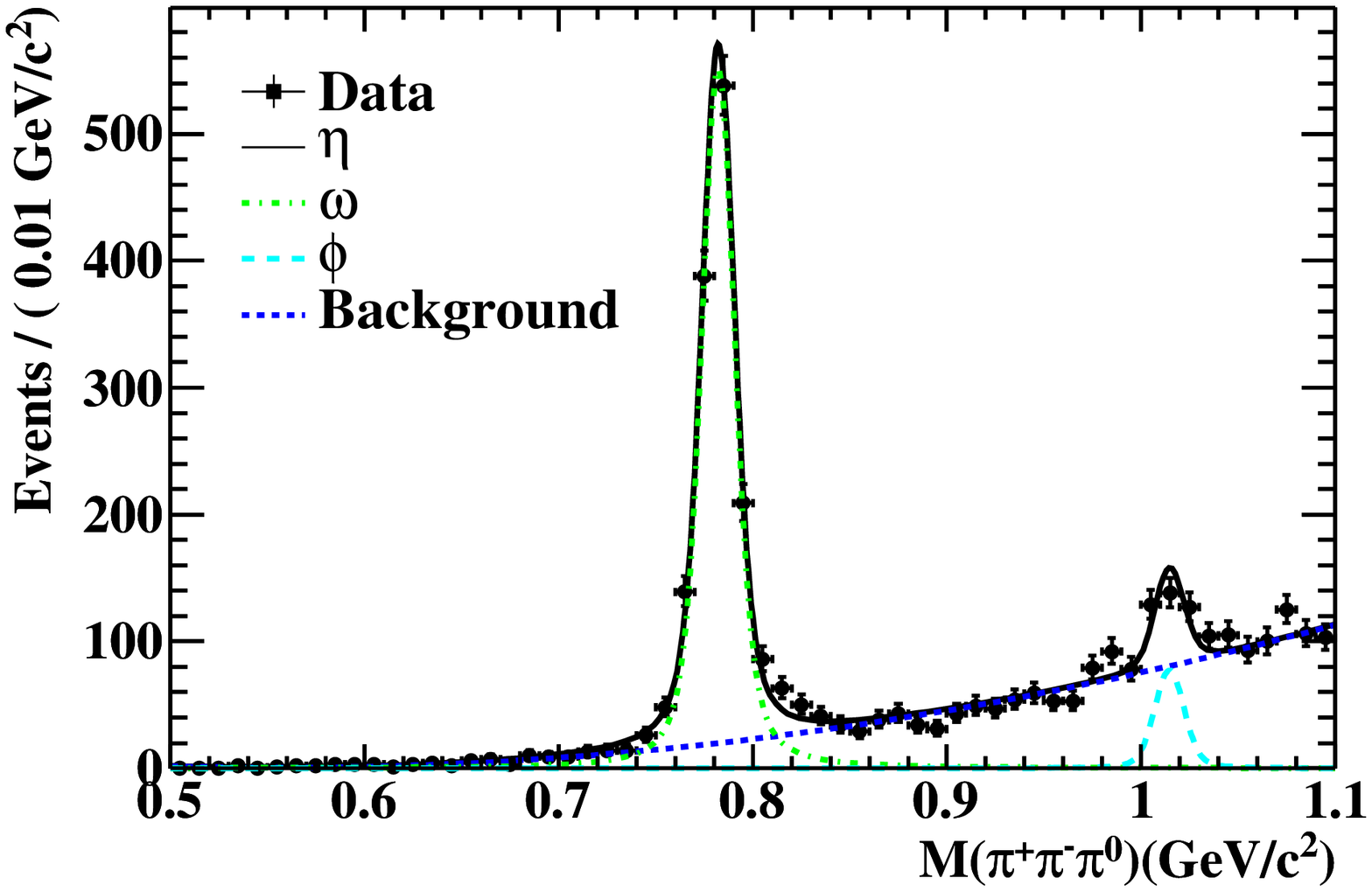}
\includegraphics[width=2.5in,height=2.0in]{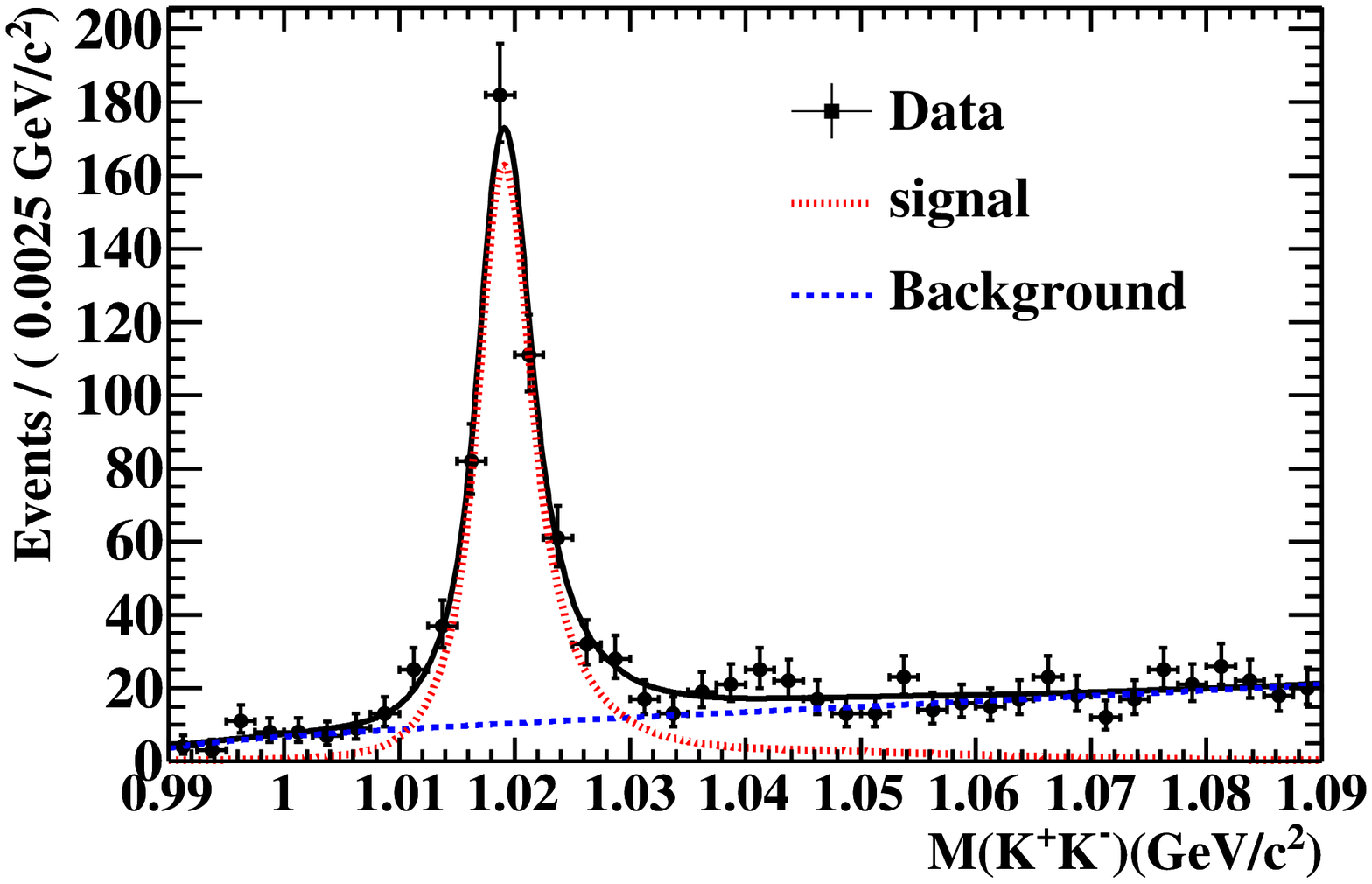}
\\
\includegraphics[width=2.5in,height=2.0in]{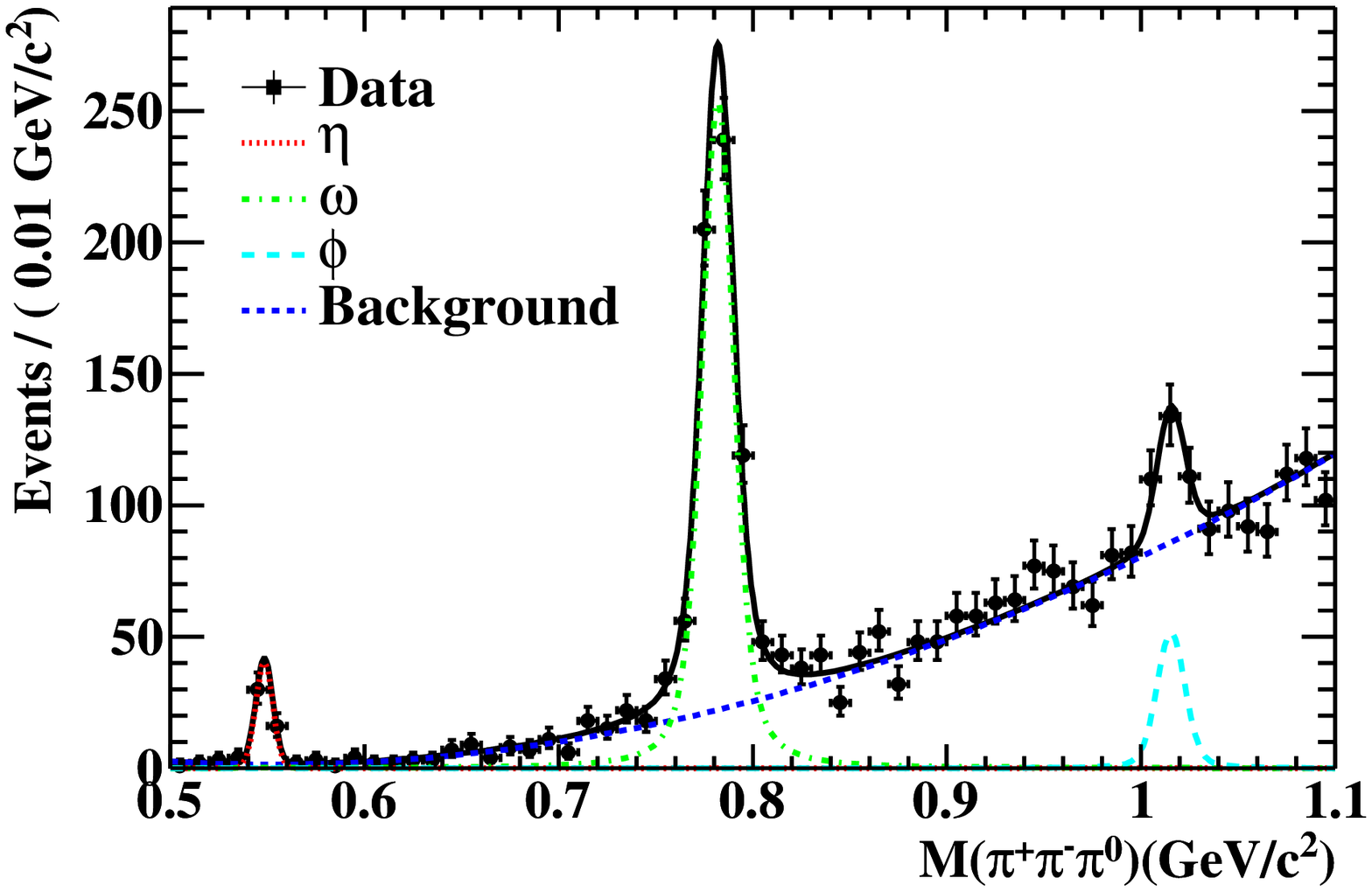}
\includegraphics[width=2.5in,height=2.0in]{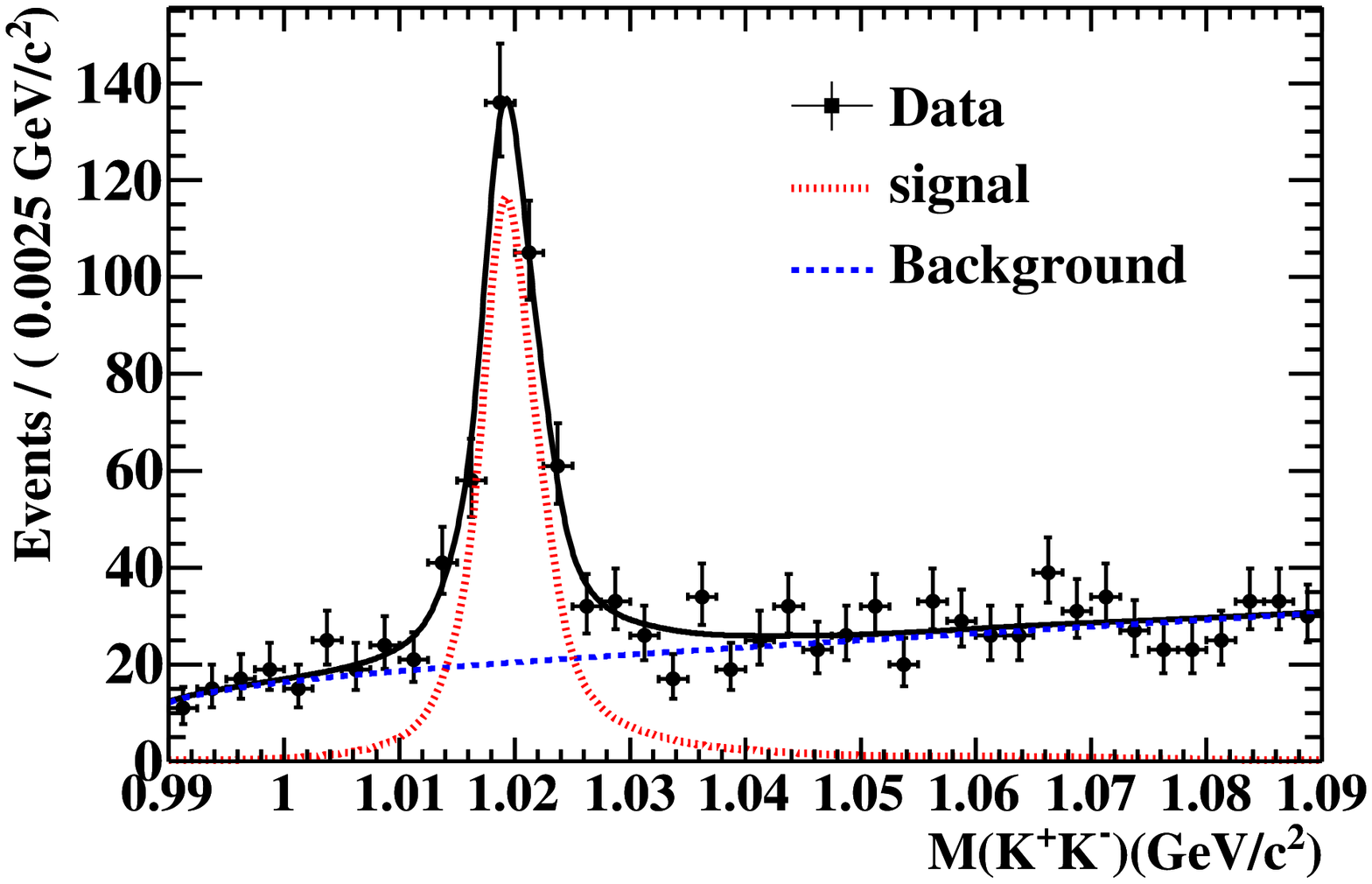}
\\
\includegraphics[width=2.5in,height=2.0in]{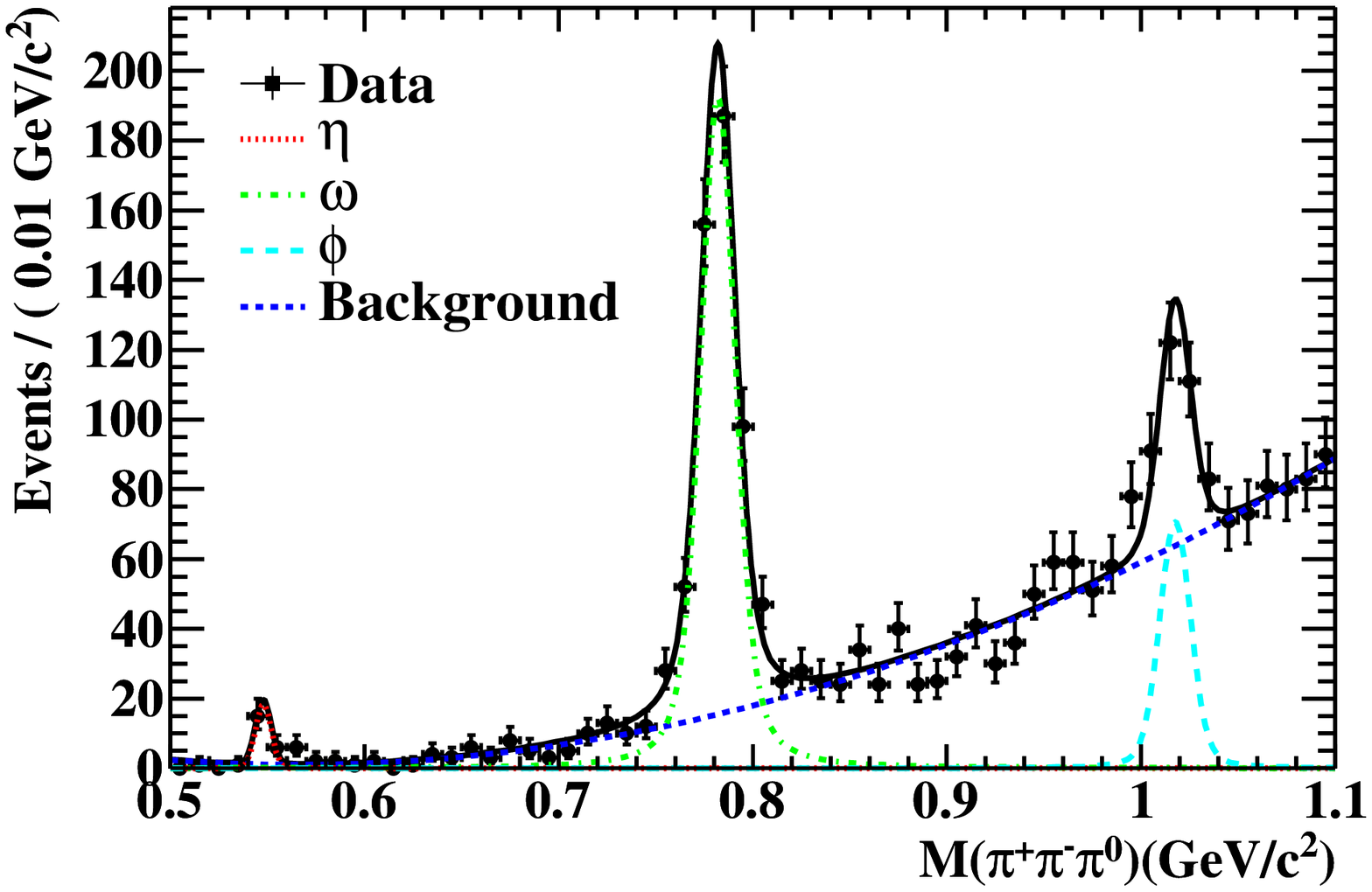}
\includegraphics[width=2.5in,height=2.0in]{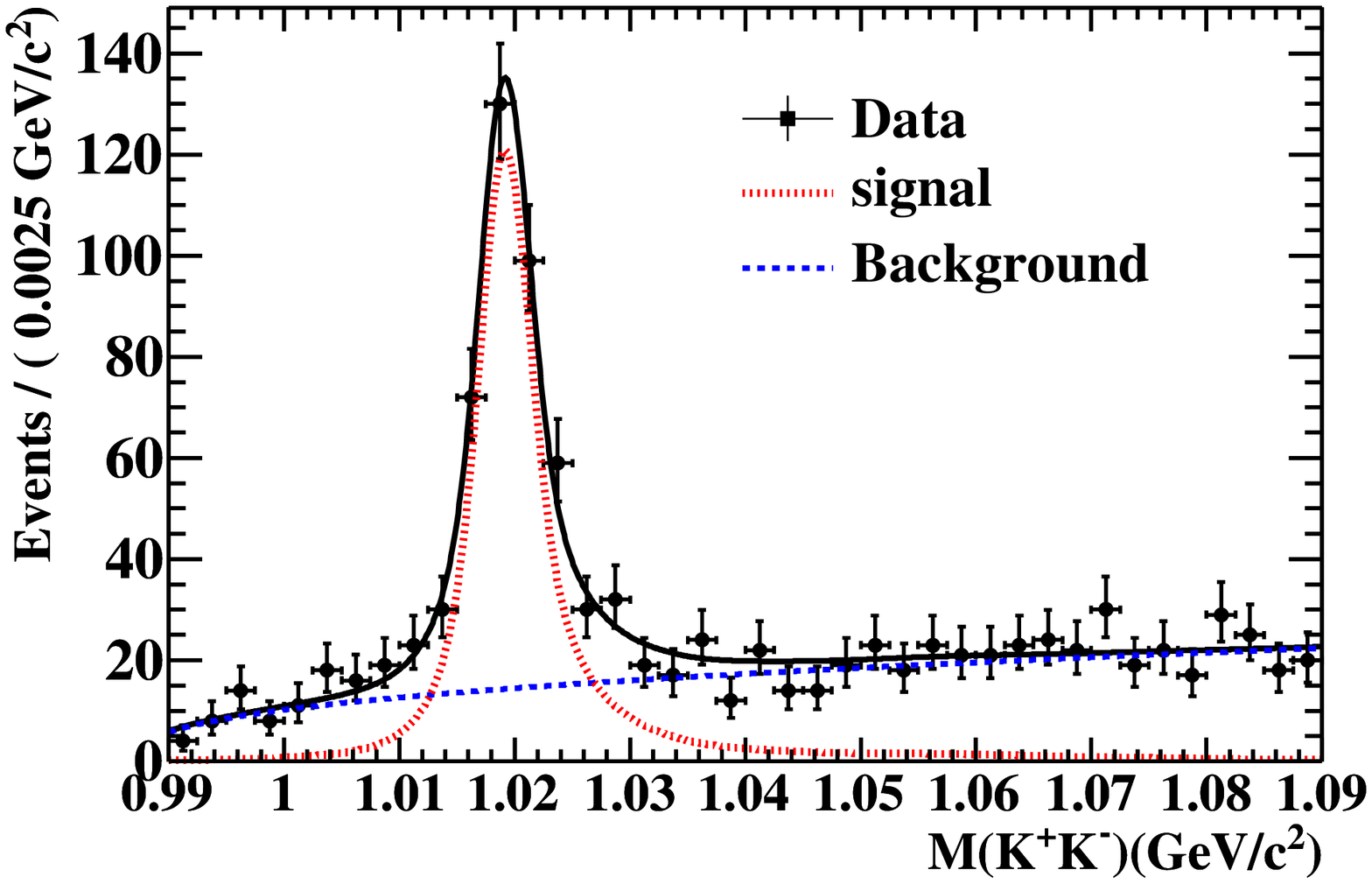}
\caption{Fits to the invariant mass spectrum of $\pp\piz$ (left
panels) and $\kk$ (right panels) in three $\ccj$ mass region after
$\ccj\to\phi\phi$, $\phi\to\kk$, $\phi\to\pp\piz$ were rejected.
From top to bottom are in $\ccz$ mass region, $\cco$ mass region,
and $\cct$ mass region, respectively.} \label{fig:fit mpipipi mkk}
\end{figure*}

In the measurement of $\BR(\ccj\to \kkppp)$, the contributions
from intermediate states with narrow resonances such as $\eta$,
$\omega$, and $\phi$ are excluded; the identification of such
resonances is similar to that used in Refs.~\cite{bes3chicjvv,
bes3gammaP}. The branching fractions of these decays $\ccj\to
\eta\kk$, $\ccj\to \omega\kk$, $\ccj\to \phi\kk$, and $\ccj\to
\phi\pp\piz$ are measured using the same data sample. The fits to
the invariant mass spectrum of $\pp\piz$ and $\kk$ in the three
$\ccj$ signal regions (as defined in Sect.~\ref{etacpp})) are
shown in Fig.~\ref{fig:fit mpipipi mkk}, and the results are
listed in Table~\ref{table:sys branch inter I}.
The first errors are statistical and the second ones are
systematic. The sources of the systematic errors are similar
to those in the measurement of $\BR(\ccj\to\kkppp)$,
as will be shown in Sect.~\ref{systematic}.
The branching fractions of $\ccj\to\omega\kk$ and
$\ccj\to\phi\pp\piz$ are the first measurement,
and those of other modes from this measurement are consistent
within errors with the known PDG values~\cite{pdg} when available.
Events from these decay modes are
removed by requiring $|m_{\pp\piz}-m_{\eta}|
> 15~\mevcc$, $|m_{\pp\piz}-m_{\omega}| > 40~\mevcc$,
$|m_{\pp\piz}-m_{\phi}| > 15~\mevcc$ and $|m_{\kk}-m_{\phi}| >
15~\mevcc$. The contribution from $\ccj\to \phi\phi$, $\phi\to
\kk$, $\phi\to \pp\piz$ events is also removed by these
requirements. The expected remaining events from these decay channels
are also listed in Table~\ref{table:sys branch inter I} and will be
subtracted from the signal yields from the best fits.

\begin{table*}[htbp]
\caption{The number of background events ($N^{\rm peak}_{\rm bkg}$) remained in
$\ccj\to \kkppp$ from modes with narrow intermediate states. The
branching fraction of $\ccj\to \phi\phi$ is taken from the BESIII
measurement~\cite{bes3chicjvv} with both statistical and
systematic errors; the other branching fractions are measured in
this analysis. We also list PDG
values~\cite{pdg} in the last column for
comparison.}\label{table:sys branch inter I}
\begin{center}
\begin{tabular}{cccc}
\hline Decay modes   & ~~~~~~~~~~$N^{\rm peak}_{\rm bkg}$~~~~~~~~   &
Branching fraction   &~~~~PDG~\cite{pdg} value~~~  \\\hline
$\ccz\to\eta\kk$       &  0   & {forbidden by $J^P$-conservation}  & $<2.3\times10^{-4}$\\
$\cco\to\eta\kk$       &  $3.81\pm0.70$   & $(3.48\pm0.57)\times10^{-4}$ & $(3.3\pm1.0)\times10^{-4}$ \\
$\cct\to\eta\kk$       &  $1.90\pm0.53$   & $(1.69\pm0.45)\times10^{-4}$ & $<3.5\times10^{-4}$  \\
\hline
$\ccz\to\omega\kk$     &  $136.87\pm10.84$ & $(1.94\pm0.06\pm0.20)\times10^{-3}$  & - \\
$\cco\to\omega\kk$     &  $69.24\pm6.20$   & $(7.82\pm0.36\pm0.84)\times10^{-4}$ & -   \\
$\cct\to\omega\kk$     &  $57.26\pm5.04$   & $(7.32\pm0.39\pm0.78)\times10^{-4}$ & -  \\
\hline
$\ccz\to\phi\kk$       &  $30.37\pm4.91$ & $(1.15\pm0.17)\times10^{-3}$   & $(0.98\pm0.25)\times10^{-3}$\\
$\cco\to\phi\kk$       &  $17.61\pm3.79$  & $(6.74\pm1.37)\times10^{-4}$ & $(4.3\pm1.6)\times10^{-4}$  \\
$\cct\to\phi\kk$       &  $30.39\pm5.27$  & $(1.14\pm0.16)\times10^{-3}$ & $(1.55\pm0.33)\times10^{-3}$  \\
\hline
$\ccz\to\phi\pp\piz$   &  $49.57\pm2.68$  & $(1.18\pm0.07\pm0.13)\times10^{-3}$ & -  \\
$\cco\to\phi\pp\piz$   &  $28.59\pm1.99$   & $(7.54\pm0.53\pm0.80)\times10^{-4}$ & -  \\
$\cct\to\phi\pp\piz$   &  $34.45\pm2.36$   & $(9.25\pm0.63\pm0.97)\times10^{-4}$ & -  \\
\hline
$\ccz\to\phi\phi$      &  $0.97\pm0.17$    & $(8.0\pm0.3\pm0.8)\times10^{-4}$ & $(8.2\pm0.8)\times10^{-4}$  \\
$\cco\to\phi\phi$      &  $0.38\pm0.08$    & $(4.4\pm0.3\pm0.5)\times10^{-4}$ & - \\
$\cct\to\phi\phi$      &  $1.03\pm0.20$    & $(10.7\pm0.3\pm1.2)\times10^{-4}$ & $(11.4\pm1.2)\times10^{-4}$ \\
\hline
\end{tabular}
\end{center}
\end{table*}


An unbinned maximum likelihood fit is applied to the invariant
mass spectrum of $\kskppp$ ($\kkppp$) to extract the numbers of
$\ccj$ events in Fig.~\ref{fig:invariant mass kkpipipi}. The
$\ccj$ signals are described by the corresponding MC simulated
signal shape convolved with a Gaussian function $G(\mu,\sigma)$ to
take into account the difference in the mass scale and the mass
resolution between data and MC simulation. The means ($\mu$) and
the standard deviations ($\sigma$) of the Gaussian functions are
floated parameters in the fit. In the generation of the $\ccz$ MC
events, the E1 radiative transition factor $E_{\gamma}^{3}$ is
included, where $E_{\gamma}$ is the energy of the radiative photon
in the $\psp$ rest frame. To damp the diverging tail due to the
$E_{\gamma}^{3}$ dependence, a damping function
$\frac{E_{0}^{2}}{E_{\gamma}E_{0}+(E_{\gamma}-E_{0})^{2}}$ used by
KEDR~\cite{kedr} is introduced, where $E_{0}=
(m_{\psp}^{2}-m_{\ccz}^{2})/2m_{\psp}$. The backgrounds are
described by a second-order Chebyshev function in both decay
modes.

The fit to the invariant mass spectrum of $\kskppp$ yields
$2837\pm 64$, $5180\pm 75$, and $4560\pm 71$ signal events for
$\ccz$, $\cco$, and $\cct$, respectively; while the fit to the
$\kkppp$ modes yields $9372\pm 130$, $12415\pm 126$, and $11366\pm
123$ events for $\ccz$, $\cco$, and $\cct$, respectively.

\subsection{\boldmath $\ccj\to \etac\pp$}
\label{etacpp}

\begin{figure*}[htbp]
\begin{center}
\includegraphics[width=2.1in,height=1.8in]{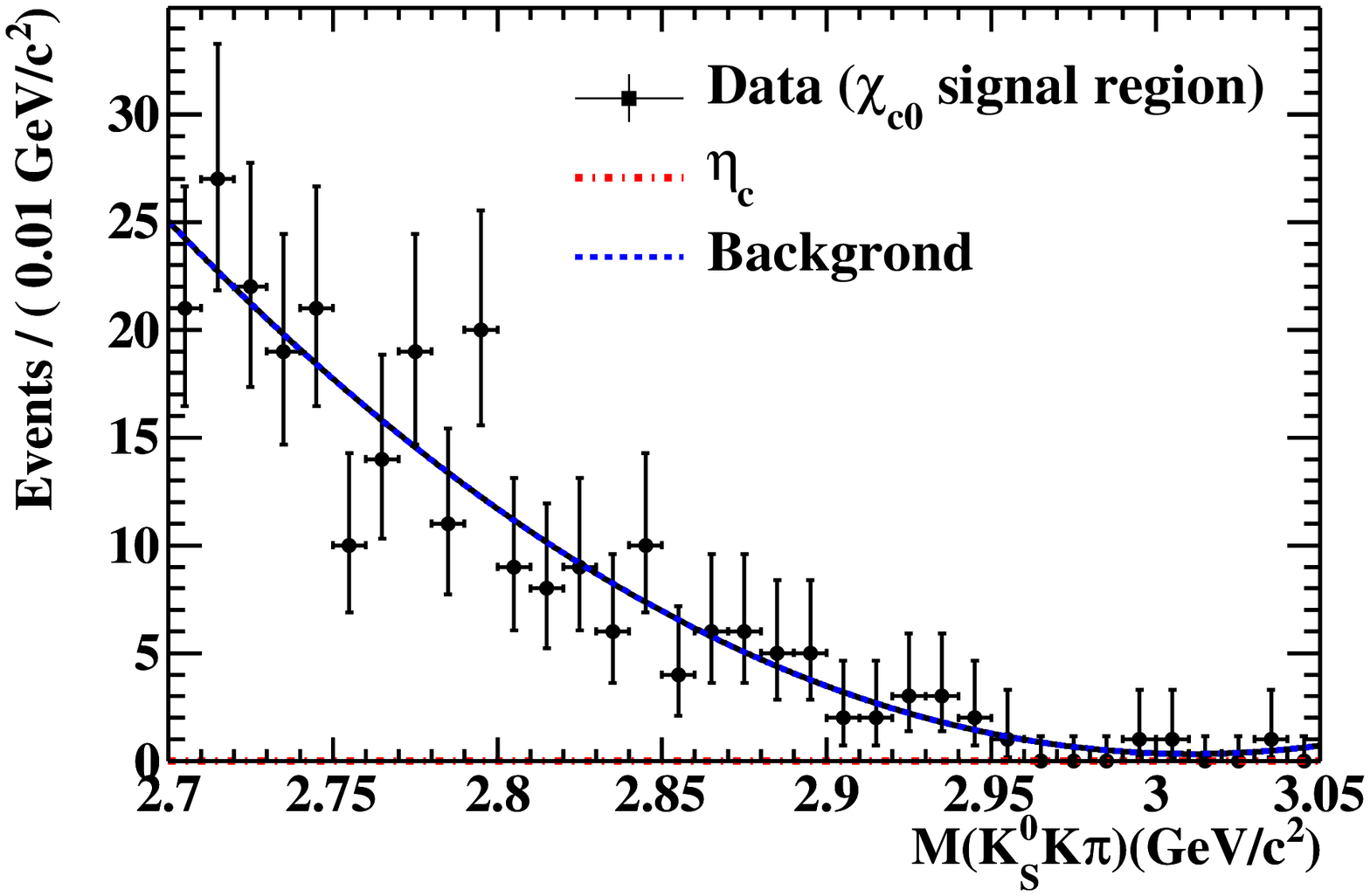}
\includegraphics[width=2.1in,height=1.8in]{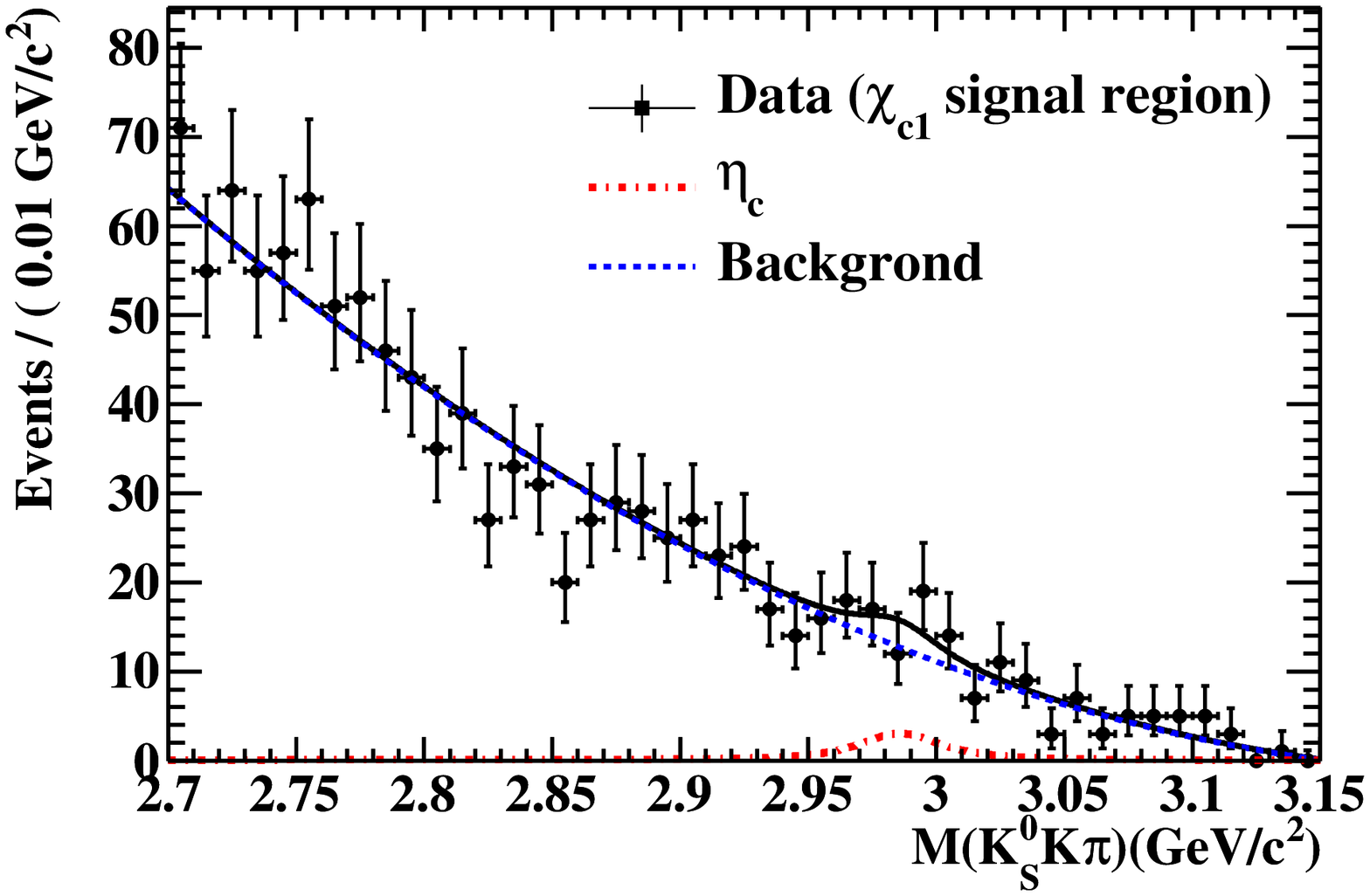}
\includegraphics[width=2.1in,height=1.8in]{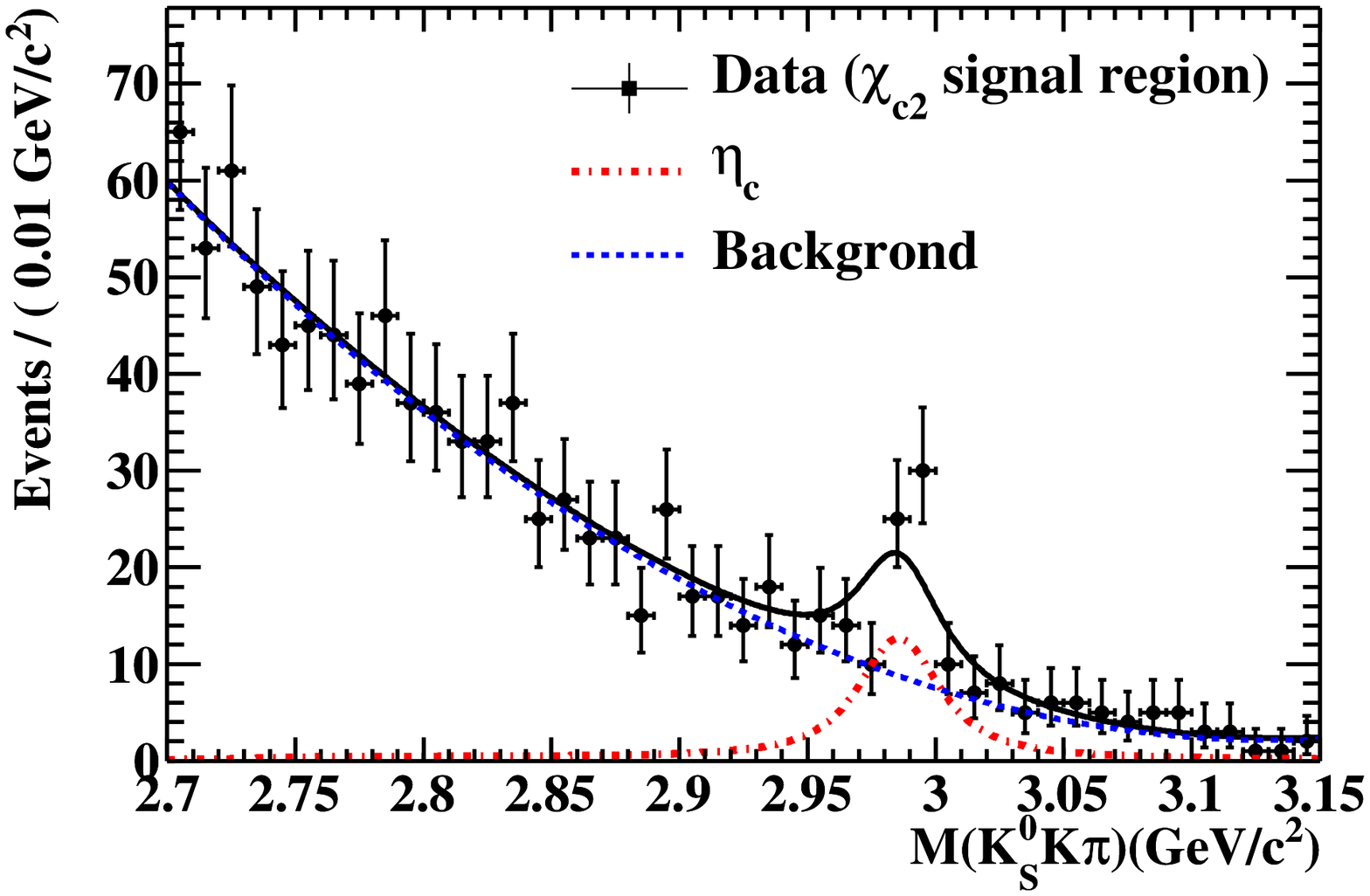}
\\
\includegraphics[width=2.1in,height=1.8in]{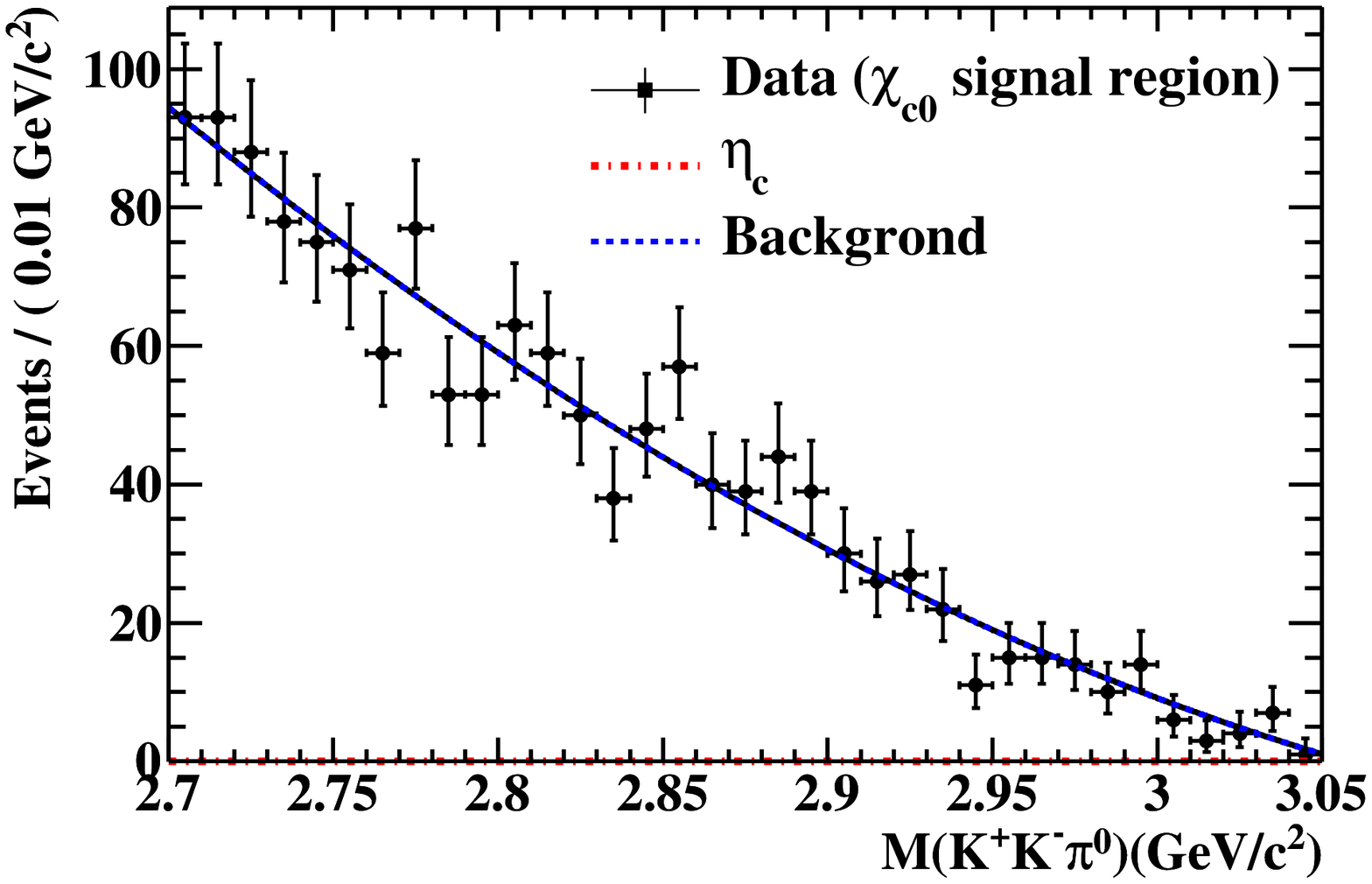}
\includegraphics[width=2.1in,height=1.8in]{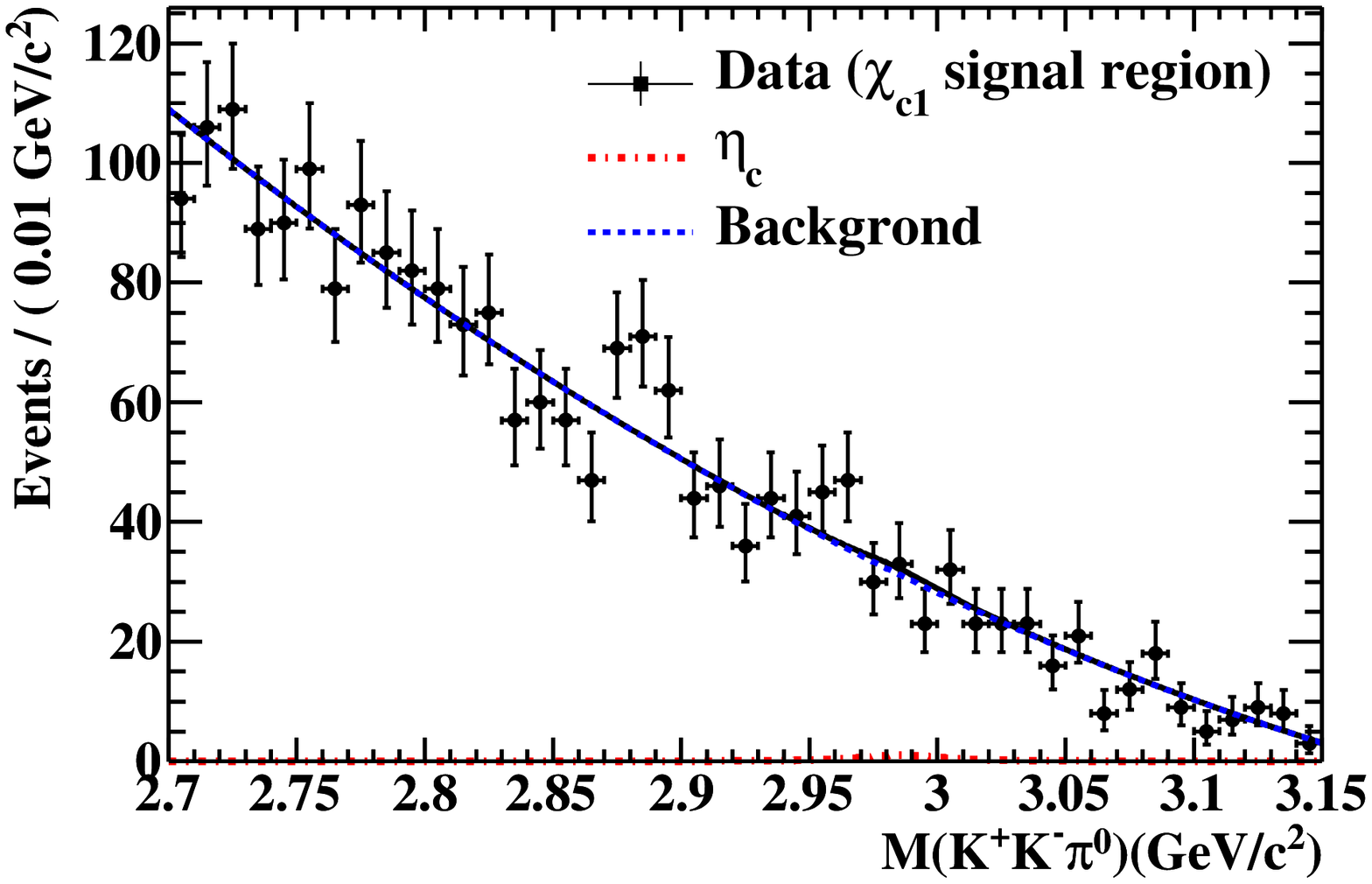}
\includegraphics[width=2.1in,height=1.8in]{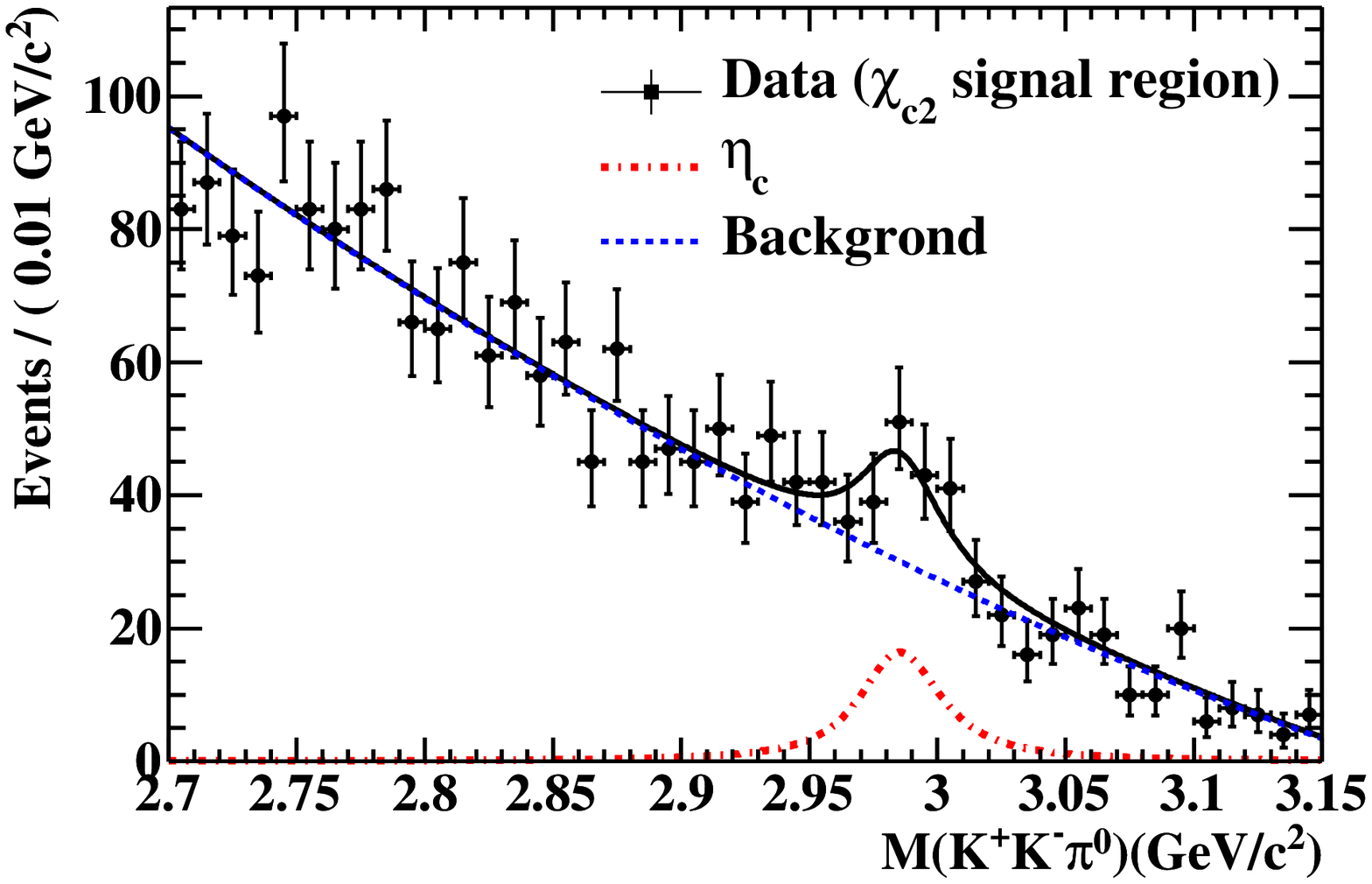}
\end{center}
\caption{Invariant mass spectra of $\kskp$ (top row) and $\kkp$
(bottom row) with $\kktp$ in $\ccz$ (left panel), $\cco$ (middle
panel), and $\cct$ (right panel) signal regions and the fit
results. Dots with error bars are data; the solid lines are the
total results from the best fits to the invariant mass spectrum.
The $\etac$ signals are shown in dash-dotted lines (in $\cct$
mass region, the contribution from the peaking background
is not removed.); the backgrounds as dashed lines.} \label{fig: invariant mass kkpi}
\end{figure*}

To study the $\ccj\to \etac\pp$ transitions, we define the signal
regions for $\ccz$, $\cco$, and $\cct$ as $3.38~\gevcc< M(\kktp)<
3.45~\gevcc$, $3.48~\gevcc< M(\kktp)< 3.54~\gevcc$, and
$3.54~\gevcc< M(\kktp)< 3.60~\gevcc$, respectively. Here, $\kktp$
is either $\kskppp$ or $\kkppp$. The invariant mass spectra of
$\kskp$ ($\kkp$) with $\kskppp$ ($\kkppp$) in the three $\ccj$
signal regions are shown in Fig.~\ref{fig: invariant mass kkpi}.
In both decay modes, there are no significant $\etac$ signal in
$\ccz$ and $\cco$ decay; the $\etac$ signal observed in $\cct$
decays is found to be mainly from $\psp\to \pp\jpsi$, $\jpsi\to
\gamma\etac$, $\etac\to \kskp$ ($\etac\to \kkp$), as described
below.

The potential backgrounds from $\psp$ decays are investigated with
the inclusive MC events. The dominant backgrounds are the
irreducible $\psp\to \gamma\ccj$, $\ccj\to \kskppp$ ($\ccj\to
\kkppp$) events. These events have the same final states as the
signal events but $\kskp$ ($\kkp$) are not from the decay of $\etac$. In
the $\cct$ signal region, there are peaking backgrounds from the decay
$\psp\to \pp\jpsi$, $\jpsi\to \gamma\etac$, $\etac\to \kskp$
($\etac\to \kkp$). The energy of the transition photon in this
decay is close to the energy of the transition photon in $\psp\to
\gamma\cct$, and the final states are the same as the signal
events. The other backgrounds are composed of dozens of decay
channels, each with a small contribution. The dominant backgrounds and
the other backgrounds contribute a smooth component in the $\etac$
mass region ($2.7~\gevcc\sim 3.2~\gevcc$), so these backgrounds are
described by a second-order Chebyshev function as shown in
Fig.~\ref{fig: invariant mass kkpi}. In the $\cct$ case, the peaking
background has the same final state and similar kinematics as the
signal events, so we use the same line-shape to describe both of
them. Data taken at $\sqrt{s}=3.65~\gev$ are used to estimate
backgrounds from the continuum process $\EE\to q\bar{q}$.  It is found
that this background is small and uniformly distributed in the full
mass region in both decay modes, so the contribution is neglected.

An unbinned maximum likelihood fit is applied to the invariant mass
spectrum of $\kskp$ ($\kkp$) to extract the number of $\etac$ events,
as shown in Fig.~\ref{fig: invariant mass kkpi}. The $\etac$ signal is
described by a MC simulated line shape with the detector resolution
included, and the resonance parameters of $\etac$ are fixed to the
latest measurement from the BESIII experiment~\cite{bes3etac}. The
background (except the peaking background in the $\cct$ signal region)
is described with a second-order Chebyshev polynomial function in both
decay modes in the three $\ccj$ signal regions.

As there is no significant $\etac$ signal in any of the three $\ccj$
states in either $\etac$ decay mode, we set upper limits on
$\BR(\ccj\to \etac\pp)$ using the probability density function (PDF)
for the expected number of signal events. In the $\ccz$ and $\cco$
signal regions, the likelihood distributions in the fitting of the
invariant mass spectra in Fig.~\ref{fig: invariant mass kkpi} are
taken as the PDFs directly. They are obtained by setting the number of
$\etac$ signal events from zero up to a very large number. In the
$\cct$ signal region, the likelihood distribution also contains the
contribution from the peaking background.  Using the known branching
fractions~\cite{pdg}, the detection efficiency from MC simulation, and
the number of $\psp$ events, the expected peaking background are
$45.7\pm11.6$ in $\kskp$ and $34.4\pm8.7$ in $\kkp$. Here, the errors
include the uncertainties in the detection efficiency and the
branching fractions. Then the PDF of signal is extracted with the PDF
of the peaking background (Gaussian distribution with mean set to the
expected number of peaking backgrounds, sigma set to its error) and
the PDF from the fit. The systematic uncertainties are considered by
smearing the PDF in each decay with a Gaussian.  The upper limit on
the number of events at the 90\% C.L. is defined as $N^{\rm up}$,
corresponding to the number of events at 90\% of the integral of the
smeared PDF. In each decay mode in the three $\ccj$ states, the
fit-related systematic errors on the number of signal yield are
estimated by using different fit ranges, different orders of the
background polynomial, and different $\etac$ line shapes with the
parameters of $\etac$ changed by one standard
deviation~\cite{bes3etac}; the maximum $N^{\rm up}$ is used in the
upper limit calculation.

\section{Systematic uncertainties}\label{systematic}

The systematic uncertainties in the measurement of $\BR(\ccj\to
K\bar{K}\pi\pi\pi)$ and $\BR(\ccj\to \etac\pp)$ are summarized in
Tables~\ref{table:sys error kk3p} and \ref{table:sys error kkp},
respectively. The systematic errors related to the MDC tracking (2\%
per track for those from IP), photon reconstruction (1\% per
photon), and $\piz$ reconstruction (1\%) are estimated with control
samples~\cite{bes3trkeff, bes3gammaP}; the errors in the branching
fractions of $\psp\to \gamma\ccj$ and $\etac\to K\bar{K}\pi$ are taken
from the PDG~\cite{pdg} and are propagated to the $\ccj$ branching
fraction measurement; and a 2\% uncertainty is taken for each decay due to
the limited statistics of the MC samples used. There is an overall 4\%
uncertainty in the branching fraction associated with the
determination of the number of $\psp$ events in our data
sample~\cite{npsp}.

\begin{table}[htbp]
\caption{Systematic errors (in \%) in $\BR(\ccj\to\kskppp)$ and
$\BR(\ccj\to\kkppp)$.}\label{table:sys error kk3p}
\begin{center}
\begin{tabular}{c|ccc|ccc}
\hline
\multirow{2}{*}{Sources}         &  \multicolumn{3}{c|}{$\kskppp$} & \multicolumn{3}{c}{$\kkppp$}       \\
                                 &  $\ccz$  &  $\cco$  &  $\cct$  &  $\ccz$        & $\cco$   & $\cct$       \\\hline
MDC tracking                     &   \multicolumn{3}{c|}{8.0}   &    \multicolumn{3}{c}{8.0}                 \\
Photon reconstruction            &   \multicolumn{3}{c|}{1.0}   &    \multicolumn{3}{c}{3.0}                 \\
MC statistics                    &  1.1  &  1.4  &  1.5  &   1.0       &    1.4      &     1.4   \\
$\ks$ reconstruction             &  1.4  &  1.6  &  1.7  &   --          &     --        &   --         \\
$\piz$ reconstruction            &  --   &   --  &   --  &   \multicolumn{3}{c}{1.0}   \\
Kinematic fit                    &  1.5  &  1.9  &  1.7  &   0.4       &    0.4      &     0.2    \\
Damping function                 &  0.5  &  0.1  &  0.1  &   0.4       &    0.1      &     0.1     \\
Intermediate states              &  1.0  &  1.0  &  1.0  &   4.0       &    4.0      &     4.0    \\
Fitting range                    &  1.0  &  0.4  &  0.2  &   0.4       &    0.4      &     0.7   \\
Background shape                 &  1.4  &  0.7  &  0.6  &   1.3       &    0.7      &     0.4    \\
$\BR(\psp\to\gamma\ccj)$         &  3.2  &  4.4  &  3.9  &   3.2       &    4.4      &     3.9   \\
Number of $\psp$ events          &   \multicolumn{3}{c|}{4.0}   &   \multicolumn{3}{c}{4.0}    \\\hline
Total                            &  10.1  &  10.5  &  10.3 & 10.9
&    11.3     &     11.2   \\\hline
\end{tabular}
\end{center}
\end{table}

\begin{table}[htbp]
\caption{Systematic errors (in \%) in $\BR(\ccj\to \etac\pp)$ in
$\etac\to\kskp$ and $\etac\to\kkp$ decay modes.}\label{table:sys
error kkp}
\begin{center}
\begin{tabular}{c|ccc|ccc}
\hline
\multirow{2}{*}{Sources}         &  \multicolumn{3}{c|}{$\etac\to \kskp$}         & \multicolumn{3}{c}{$\etac\to \kkp$}       \\
                                 & $\ccz$  & $\cco$  & $\cct$  & $\ccz$ & $\cco$ & $\cct$ \\\hline
MDC tracking                     &    \multicolumn{3}{c|}{8.0}                    &    \multicolumn{3}{c}{8.0}                 \\
Photon reconstruction            &    \multicolumn{3}{c|}{1.0}                    &    \multicolumn{3}{c}{3.0}                 \\
MC statistics                    &    1.8       &     1.5      &     1.6      &    1.8      &     1.5     &   1.7       \\
$\ks$ reconstruction             &    2.4       &     2.3      &     2.2      &    --         &      --       &    --         \\
$\piz$ reconstruction            &    --        &     --       &     --       &     \multicolumn{3}{c}{1.0}          \\
Kinematic fit                    &    1.5       &     1.7      &     1.8      &    0.9      &     0.2     &   0.4       \\
$\BR(\psp\to\gamma\ccj)$ &    3.2       &     4.4      &     3.9      &    3.2      &     4.4     &   3.9       \\
$\BR(\etac\to K\bar{K}\pi)$ &    \multicolumn{3}{c|}{8.4}                   &    \multicolumn{3}{c}{8.4}                \\
Number of $\psp$ events          &    \multicolumn{3}{c|}{4.0}                    &    \multicolumn{3}{c}{4.0}                 \\\hline
Total                            &    13.2      &     13.5     &
13.4     &    13.2     &     13.5    &   13.4
\\\hline
\end{tabular}
\end{center}
\end{table}

\subsection{\boldmath $\ks$ reconstruction}

The uncertainty in the $\ks$ reconstruction arises from three
parts: the geometric acceptance, the tracking efficiency, and the
efficiency of $\ks$ selection. The first part is estimated using
MC simulation, and the other two are studied using
$\jpsi\to K^{*}{}^{\pm}K^{\mp}$, $K^{*}{}^{\pm}\to
K^{0}\pi^{\pm}$. By selecting a pair of $K^{\pm}\pi^{\mp}$, the
recoil mass spectrum shows a clear $K^0$ signal. The efficiency of
$\ks$ reconstruction is calculated with $\frac{n_{1}} {\BR(\ks\to
\pp)\times (n_{1}+n_{2})/2}$, where $n_{1}$ is the number of
$K^{0}$ obtained from a fit to the $K^{\pm}\pi^{\mp}$ recoiling mass
when there is a $\ks$ reconstructed in the recoil side that satisfies
the $\ks$ selection, and $n_{2}$ is the number of $K^{0}$ from fitting
to the $K^{\pm}\pi^{\mp}$ recoiling mass spectrum when no $\ks$
candidate satisfies the $\ks$ selection. The difference in the
efficiency of $\ks$ reconstruction ($\eff^{\rm data}/\eff^{\rm MC}-1$)
as a function of $\ks$ momentum is shown in Fig.~\ref{fig:fit ks
uncertainty}. The difference in the $\ks$ reconstruction between data
and MC simulation is fitted with a linear function of the $\ks$
momentum as shown in Fig.~\ref{fig:fit ks uncertainty} together with
the $\pm 1\sigma$ envelops. Since the difference between data and MC
is significant, we do a correction to the signal MC according to the
momentum of $\ks$, and the uncertainty of this correction is taken as
the systematic error.

The systematic errors in $\BR(\ccj\to \kskppp)$ are found to be
1.4\%, 1.6\%, and 1.7\%, for $J=0$, 1, and 2, respectively; while for
$\BR(\ccj\to \etac\pp)$, they are 2.4\%, 2.3\%, and 2.2\% for $\ccz$,
$\cco$, and $\cct$, respectively.

\begin{figure}[htbp]
\centerline{\psfig{file=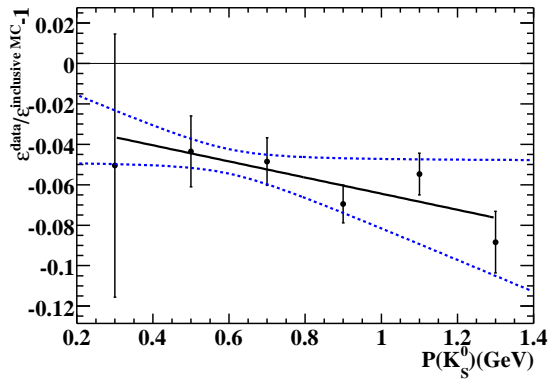,width=0.45\textwidth}}
\caption{The difference in the $\ks$ reconstruction efficiency
between data and MC simulation (points with error bars), together
with the fit to the difference with a linear function of momentum.
The solid line is from the best fit and the dashed lines are the
$\pm 1\sigma$ envelopes of the best fit.} \label{fig:fit ks
uncertainty}
\end{figure}

\subsection{Kinematic fit}

In the MC simulation, the model is much simpler than the real detector
performance, and this results in differences between data and MC
simulation in the track parameters of photons and charged tracks. The
simulation of the photon has been checked in another
analysis~\cite{npsp}, which shows good agreement between data and MC
simulation. For the charged tracks, careful comparisons with purely
selected data samples indicate that the MC simulates the momentum and
angular resolutions significantly better than those in data, while the
error matrix elements agree well between data and MC simulation. This
results in a much narrower $\chi^2_{\rm 4C}$ distribution in MC than
in data, and introduces a bias in the efficiency estimation.  We
correct the track helix parameters of MC simulation to reduce the
difference between data and MC simulation.

We use $\jpsi\to \phi f_{0}(980)$, $\phi\to \kk$, $f_{0}(980)\to
\pp$ as a control sample to study the difference on the helix
parameters of charged tracks between data and MC, as this channel
has a large production rate, very low background, and has both pions
and kaons. We find that the pull distributions of data are wider
than MC simulation and the peak positions are shifted. These
obvious differences between data and MC suggest wrong track
parameters have been set in MC simulation. The helix parameters of
each track in the MC simulation are enlarged by smearing with a
Gaussian function $G(\mu, \sigma)= G((\mu_{i}^{\rm data}-\mu_{i}^{\rm MC})
\times V_{ii}$, $ \sqrt{(\sigma_{i}^{\rm data}/\sigma_{i}^{\rm MC})^{2}-1}
\times V_{ii})$, where $i=\{d\rho,\phi_{0},\kappa,dz,tg\lambda\}$
is the $i$-th helix parameter of the track and $V$ is the
corresponding covariance matrix. Here
$d\rho$ is the distance from the pivot to
the orbit in the $x$-$y$ plane, $\phi_{0}$ is the
azimuthal angle specifies the pivot
with respect to the helix center, $\kappa$ is the reciprocal
of the transverse momentum, $dz$ is the distance of the helix
from the pivot to the orbit in the $z$ direction,
and $tg\lambda$ is the slope of the track.
The correction factors
$\mu_{i}^{\rm data}$, $\mu_{i}^{\rm MC}$, $\sigma_{i}^{\rm data}$, and
$\sigma_{i}^{\rm MC}$ are the means and resolutions of the pull
distributions of the data and MC obtained from control samples.

The correction factors are listed in Table~\ref{table:sys
correction factors}. If the correction is perfect, $\chi^{2}_{\rm
4C}$ distributions of data and MC simulation will be consistent with
each other; however, from the comparisons of many final states, we
find that the agreement between data and MC simulation does
improve significantly but differences still exist. This indicates
that the effect is from multiple sources, and our procedure cannot
solve all the problems. In our analysis, we take the efficiency
from the track-parameter-corrected MC samples as the nominal
value, and take half of the difference between MC samples before
and after the correction as the systematic error from the kinematic
fitting. This is a very conservative estimation. The comparison of
$\chi^{2}_{\rm 4C}$ distributions between data and MC simulation
before and after the track-parameter-correction are shown in
Figs.~\ref{fig:sys compare chisq kskpi} and~\ref{fig:sys compare
chisq kkpi} for $\ccj\to \kskppp$ and $\kkppp$, respectively.

\begin{table*}[htbp]
\caption{Correction factors extracted from pull distributions
using a control sample of $\jpsi\to \phi f_{0}(980)$, $\phi\to \kk$,
$f_{0}(980)\to \pp$.}\label{table:sys correction factors}
\begin{center}
\begin{tabular}{c|cc|cc|cc}
\hline
\multirow{2}{*}{}         &     \multicolumn{2}{c|}{$\phi_{0}$}
& \multicolumn{2}{c|}{$\kappa$}   & \multicolumn{2}{c}{$tg\lambda$} \\
         &$\mu^{\rm data}-\mu^{\rm MC}$&$\sigma^{\rm data}/\sigma^{\rm MC}$
         &$\mu^{\rm data}-\mu^{\rm MC}$&$\sigma^{\rm data}/\sigma^{\rm MC}$
         &$\mu^{\rm data}-\mu^{\rm MC}$&$\sigma^{\rm data}/\sigma^{\rm MC}$\\\hline
$K^{+}$  & $-0.04\pm0.03$  &     $1.19\pm0.02$
         & $-0.24\pm0.03$  &     $1.28\pm0.02$
         & $-0.38\pm0.01$  &        $1.25\pm0.02$      \\
$K^{-}$  & $ 0.06\pm0.03$  &     $1.21\pm0.02$
         & $ 0.25\pm0.03$  &     $1.25\pm0.02$
         & $-0.36\pm0.01$  &        $1.21\pm0.02$      \\
$\pi^{+}$& $-0.06\pm0.03$  &     $1.25\pm0.02$
         & $-0.10\pm0.03$  &     $1.31\pm0.02$
         & $-0.36\pm0.01$  &        $1.25\pm0.02$      \\
$\pi^{-}$& $-0.02\pm0.03$  &     $1.23\pm0.02$
         & $ 0.10\pm0.03$  &     $1.27\pm0.02$
         & $-0.36\pm0.01$  &     $1.21\pm0.02$
\\\hline
\end{tabular}
\end{center}
\end{table*}

\begin{figure*}[htbp]
\includegraphics[width=2.5in,height=2.0in]{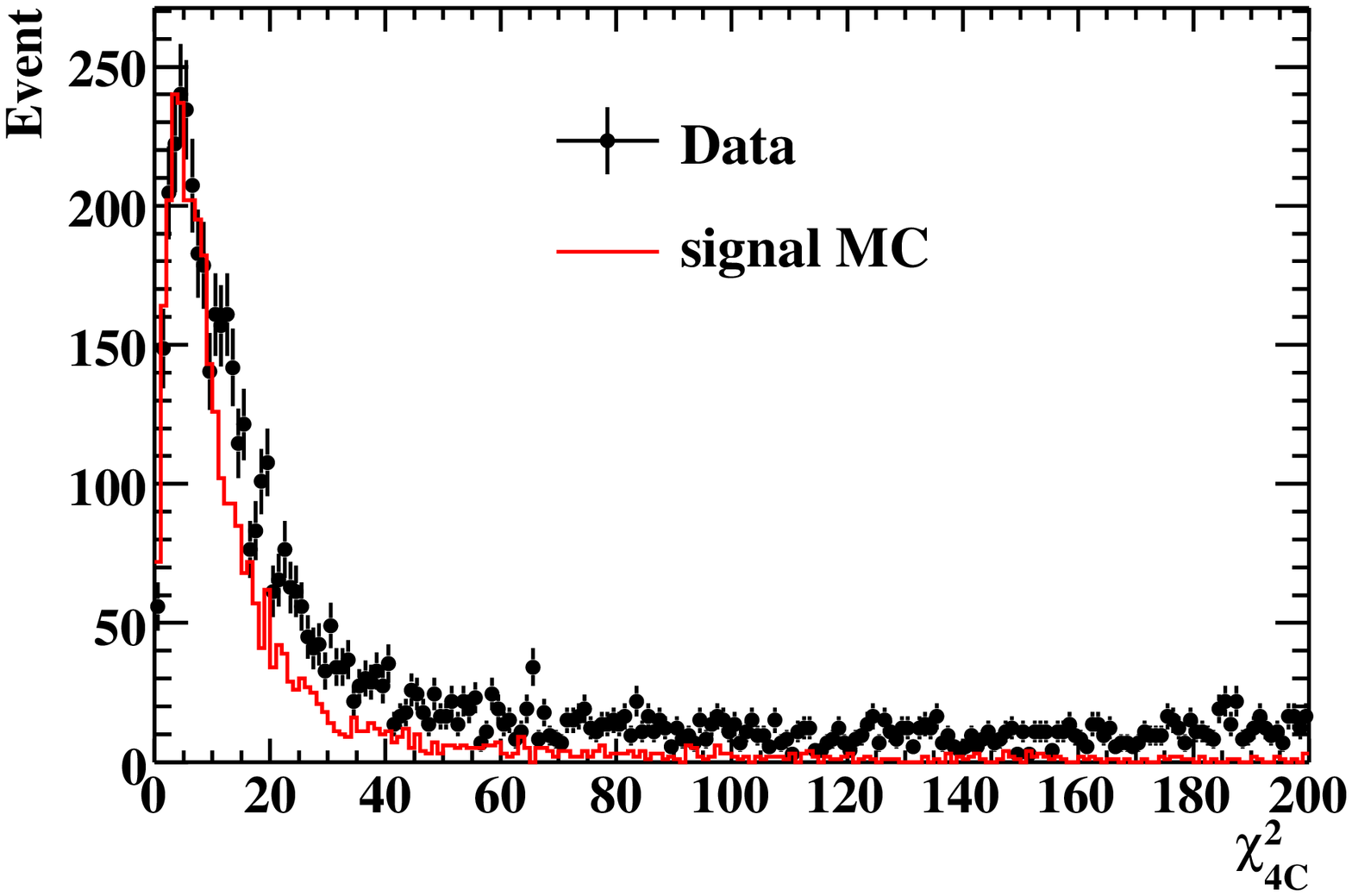}
\includegraphics[width=2.5in,height=2.0in]{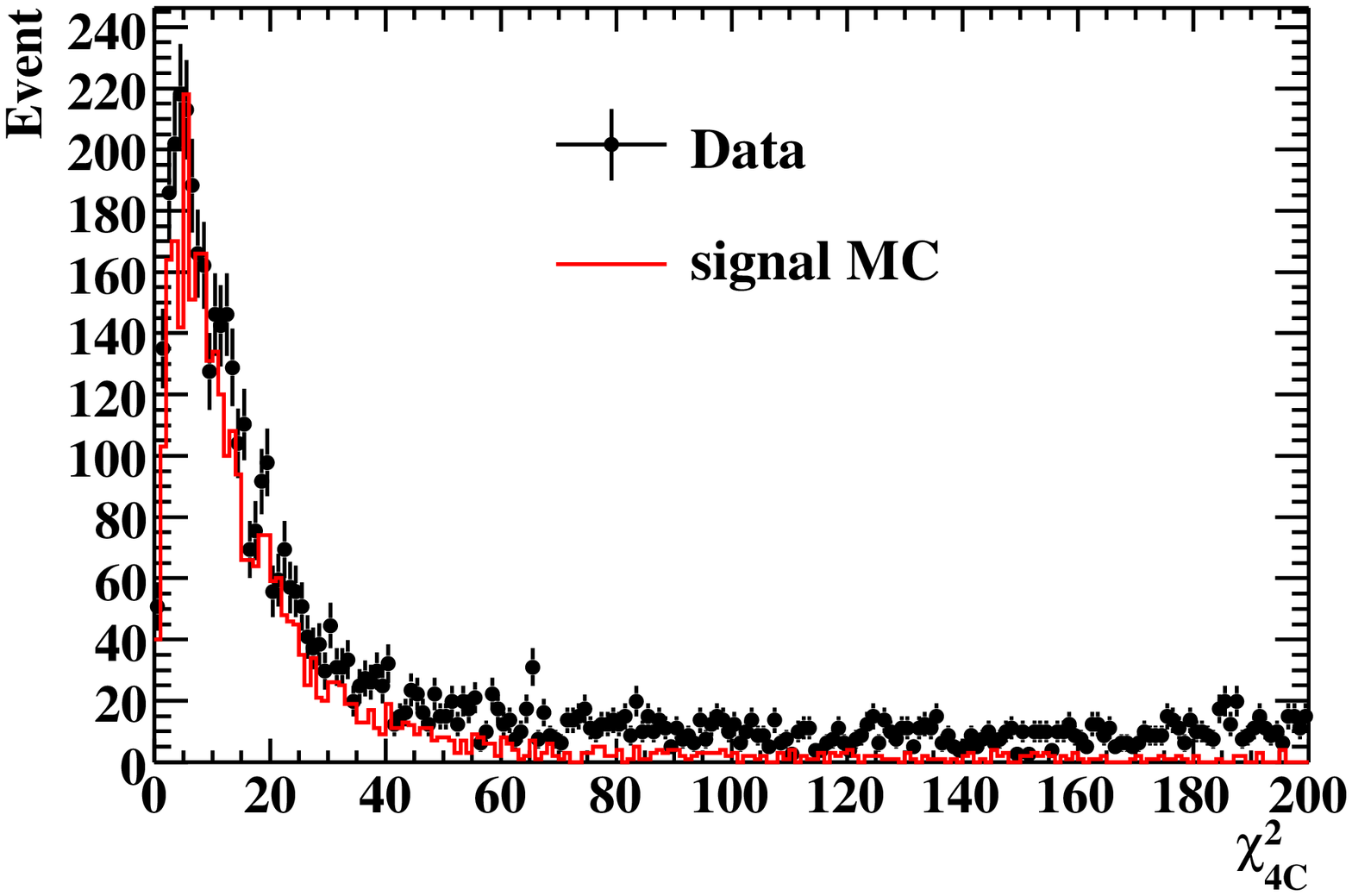}
\\
\includegraphics[width=2.5in,height=2.0in]{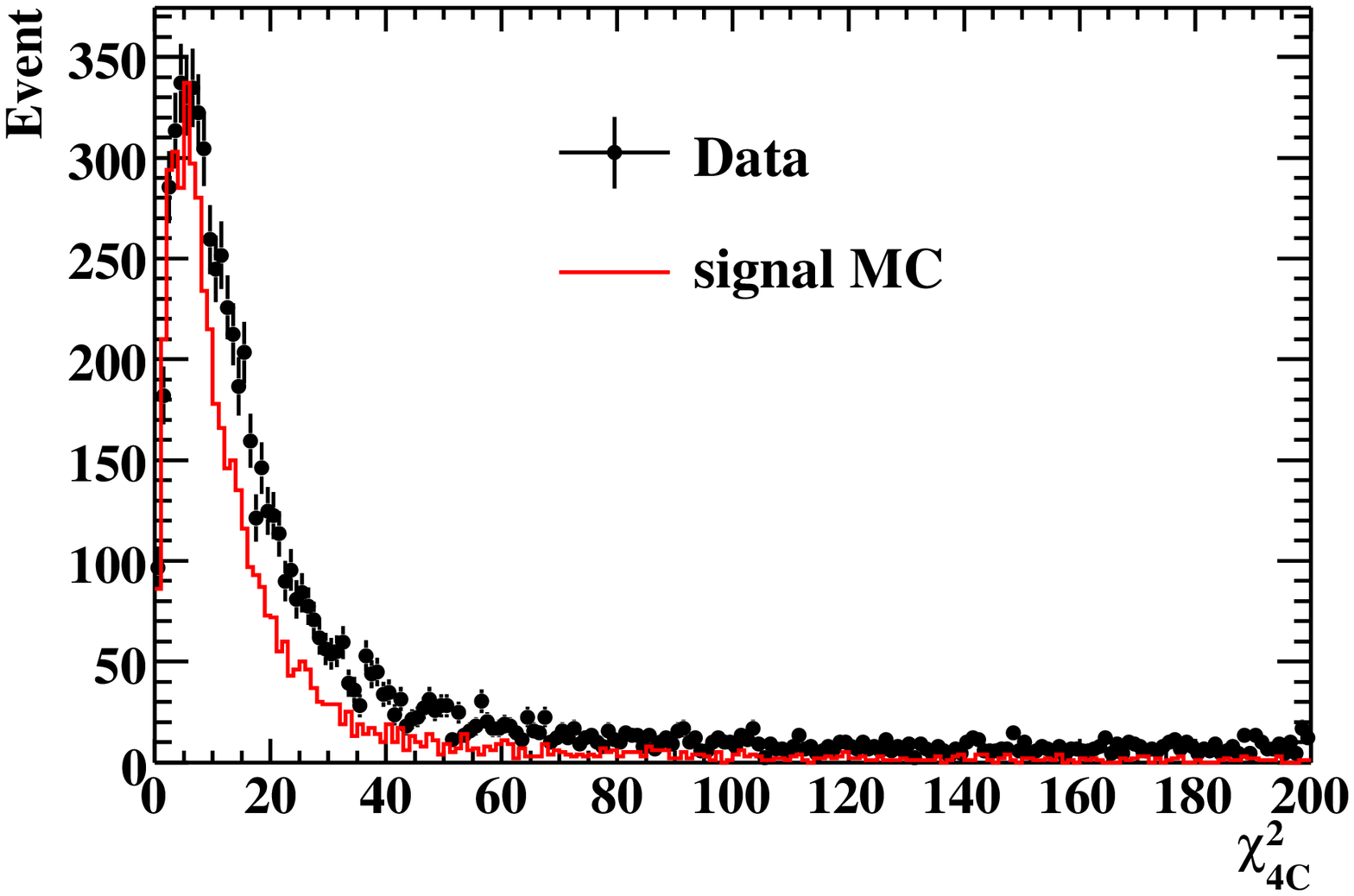}
\includegraphics[width=2.5in,height=2.0in]{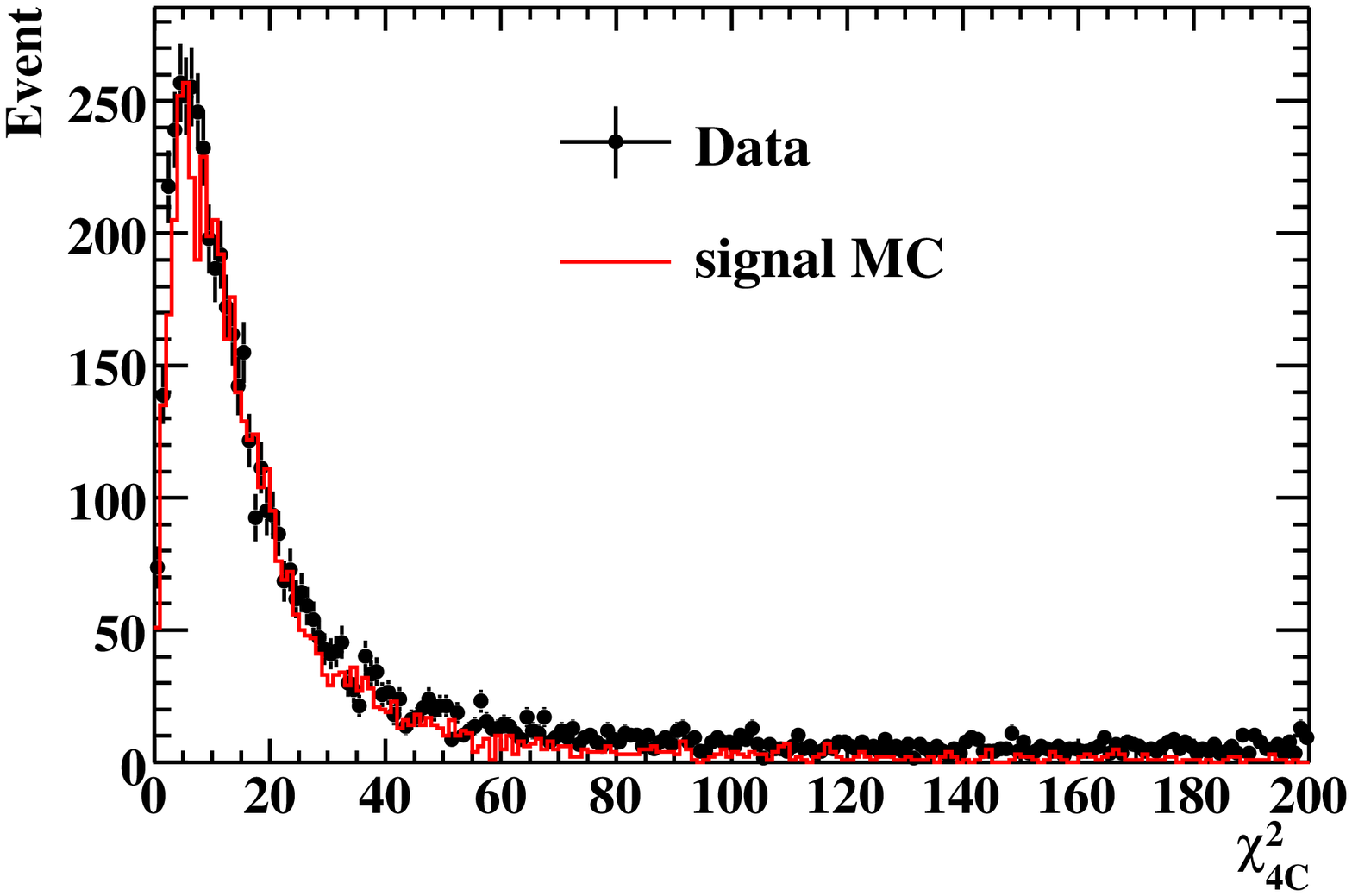}
\\
\includegraphics[width=2.5in,height=2.0in]{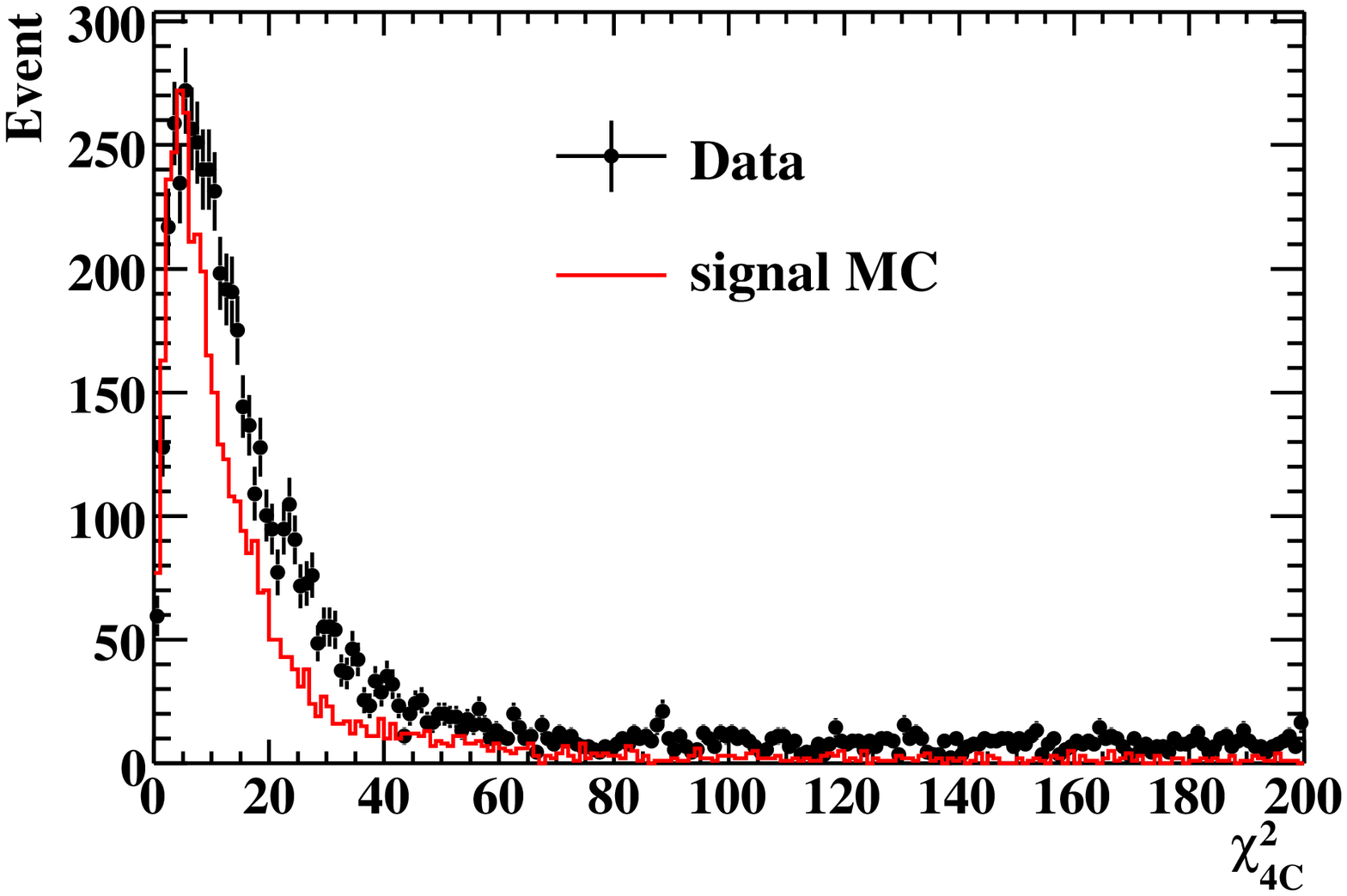}
\includegraphics[width=2.5in,height=2.0in]{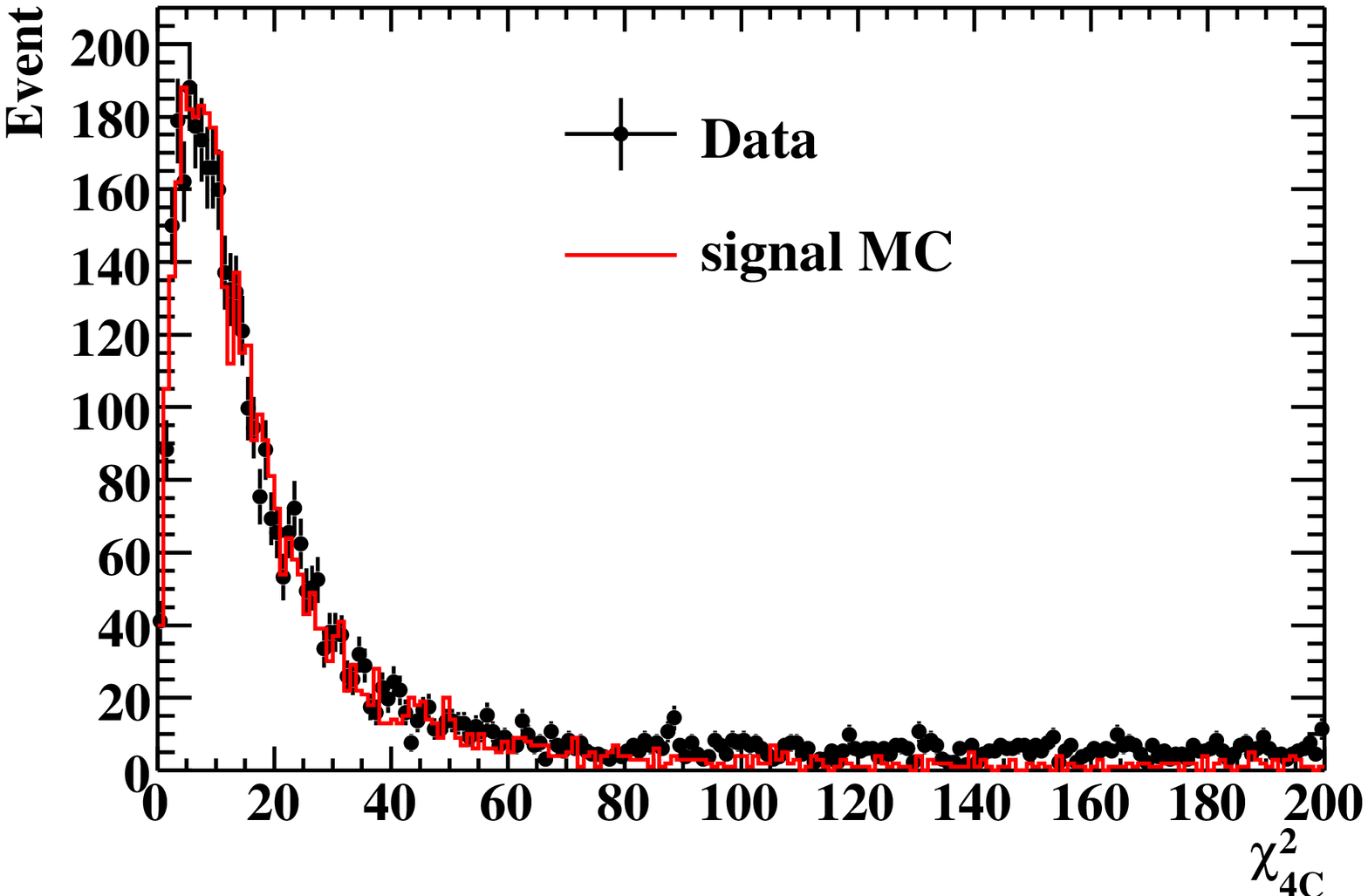}
\caption{Comparison of $\chi^{2}_{\rm 4C}$ between signal MC and
data for $\psp\to \gamma\ccj$, $\ccj\to \kskppp$. The points with
error bars are data, and the solid lines are MC simulation. Left
panel: signal MC without track-parameter-correction; Right panel:
signal MC after track-parameter-correction. From top to bottom are
$\ccz$, $\cco$, and $\cct$, respectively.} \label{fig:sys compare
chisq kskpi}
\end{figure*}

\begin{figure*}[htbp]
\includegraphics[width=2.5in,height=2.0in]{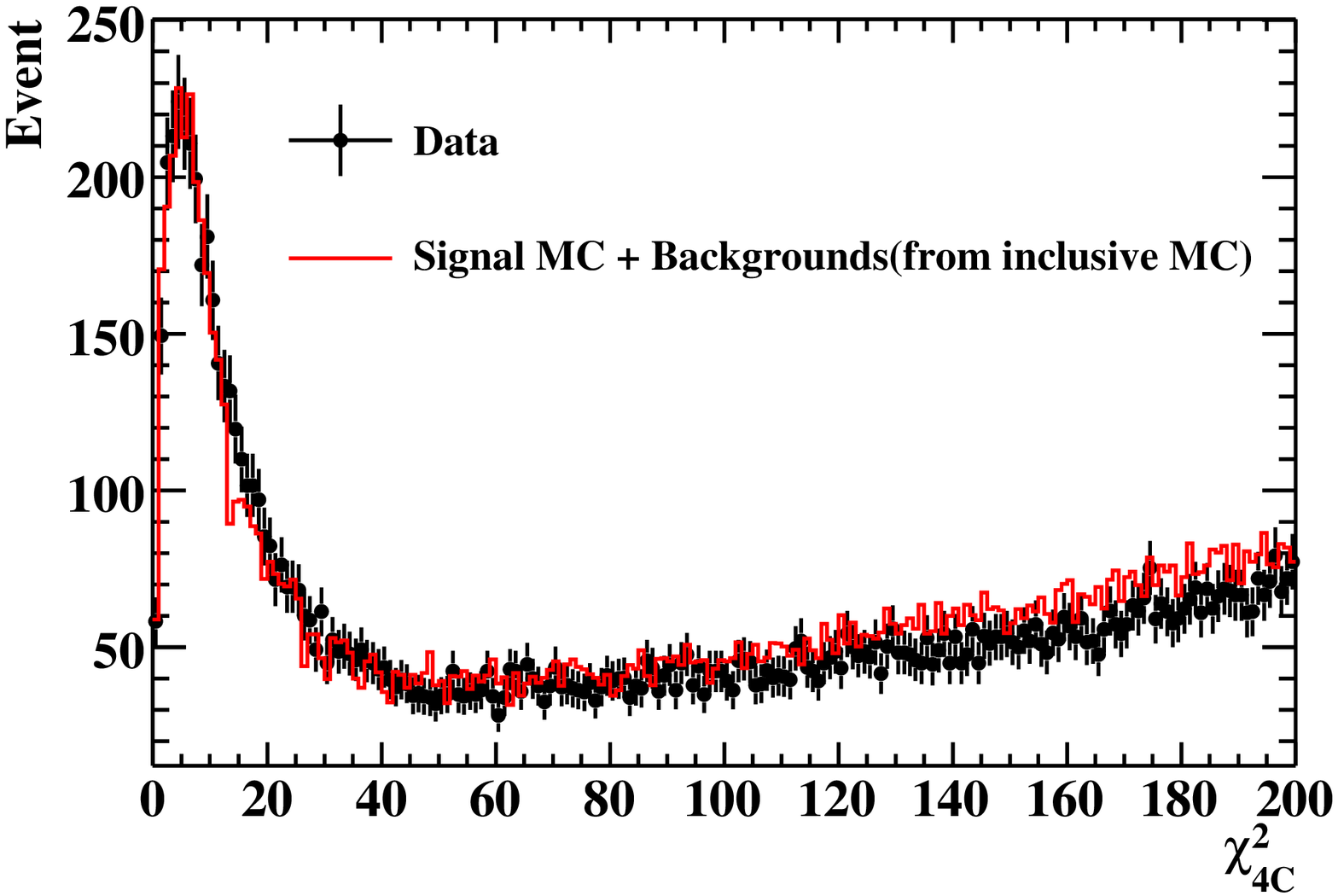}
\includegraphics[width=2.5in,height=2.0in]{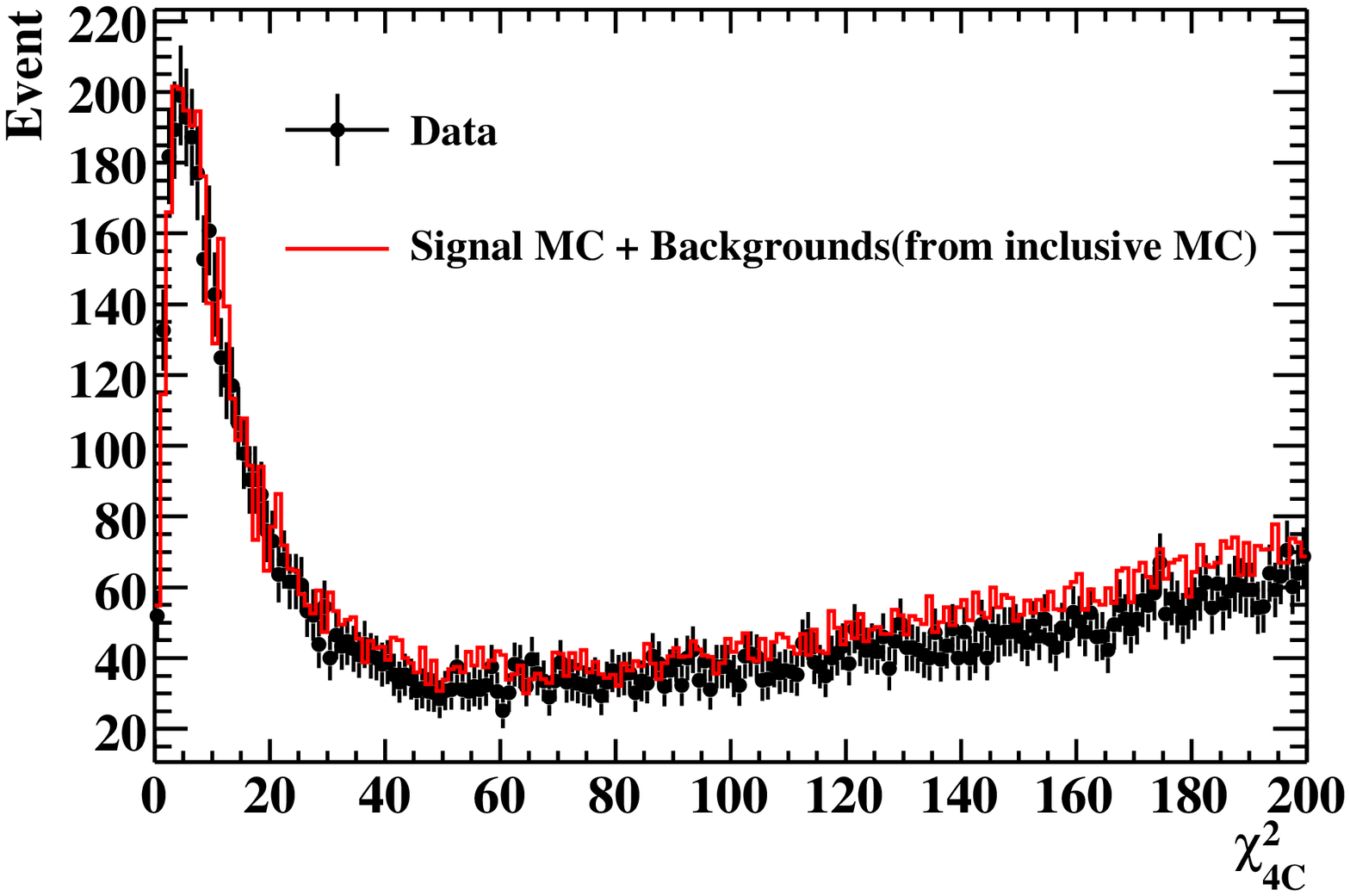}
\\
\includegraphics[width=2.5in,height=2.0in]{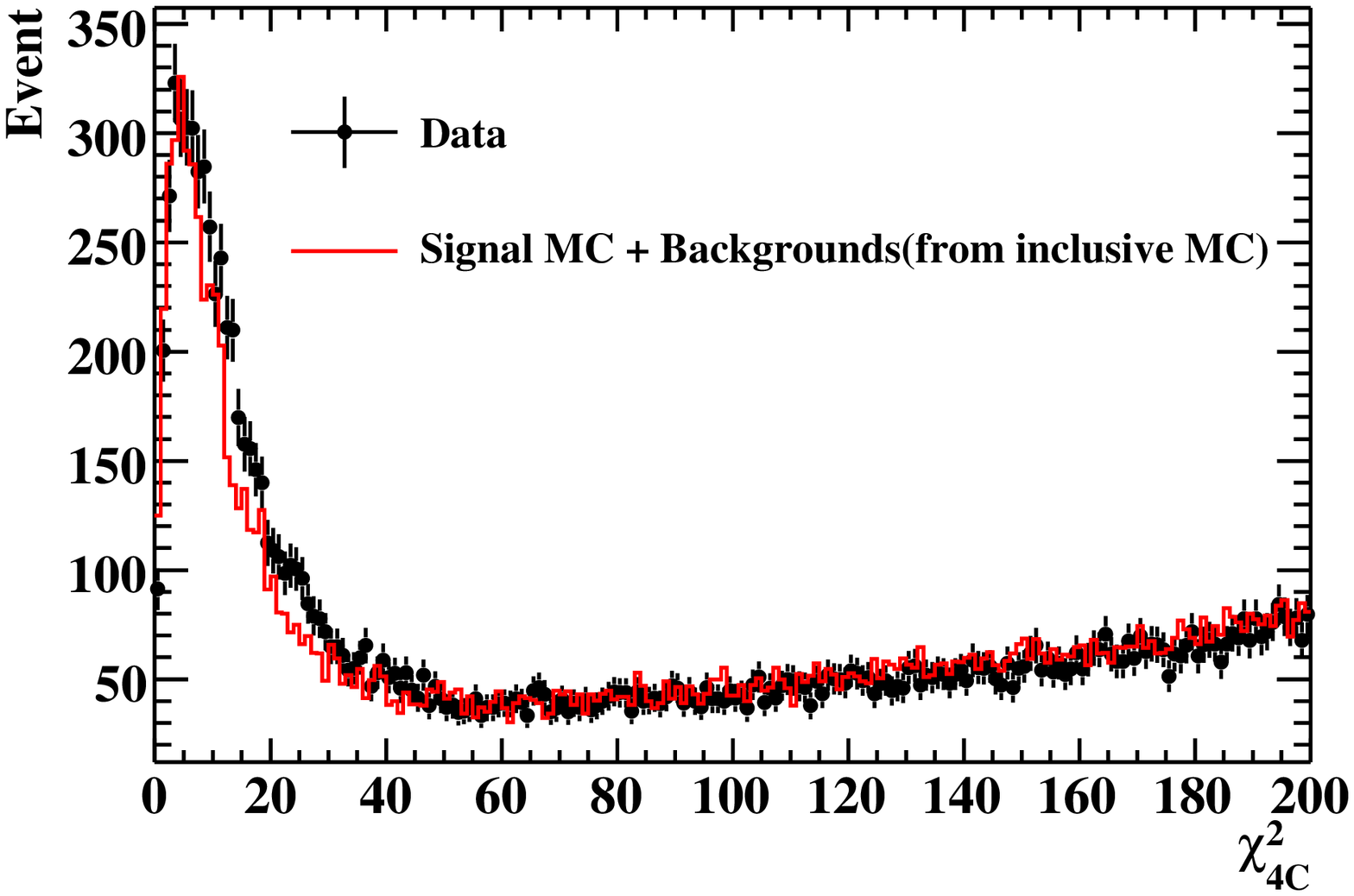}
\includegraphics[width=2.5in,height=2.0in]{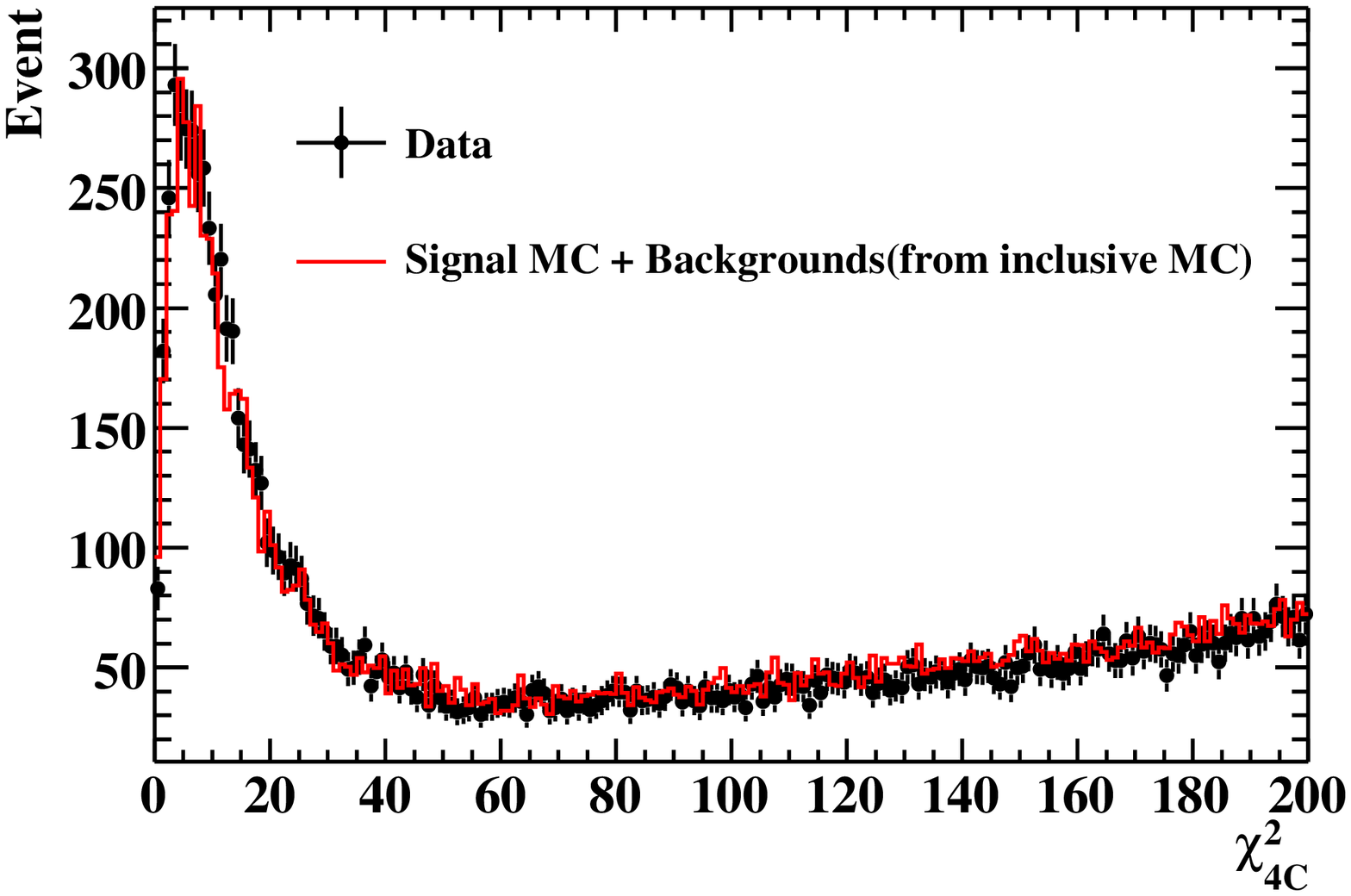}
\\
\includegraphics[width=2.5in,height=2.0in]{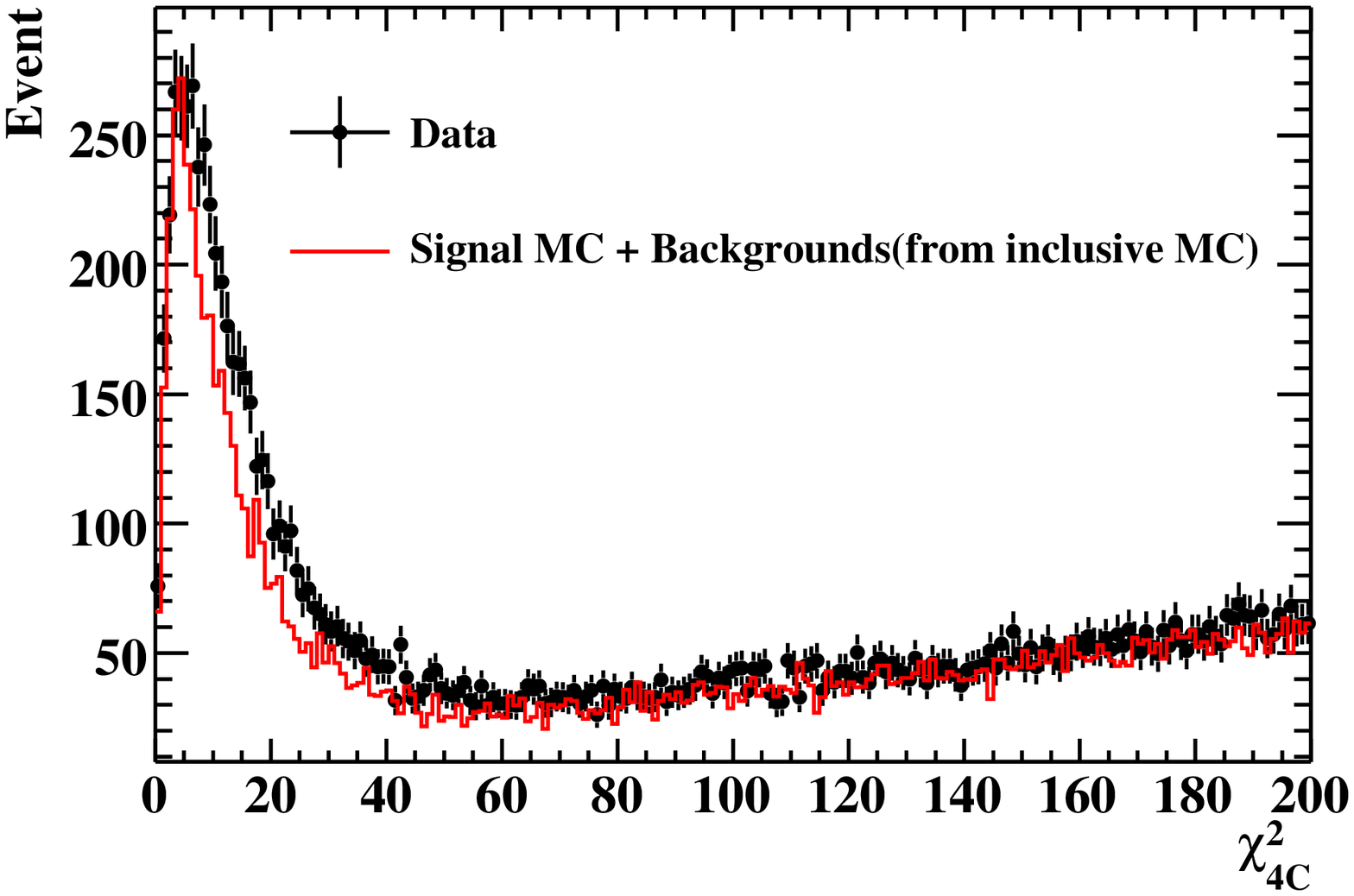}
\includegraphics[width=2.5in,height=2.0in]{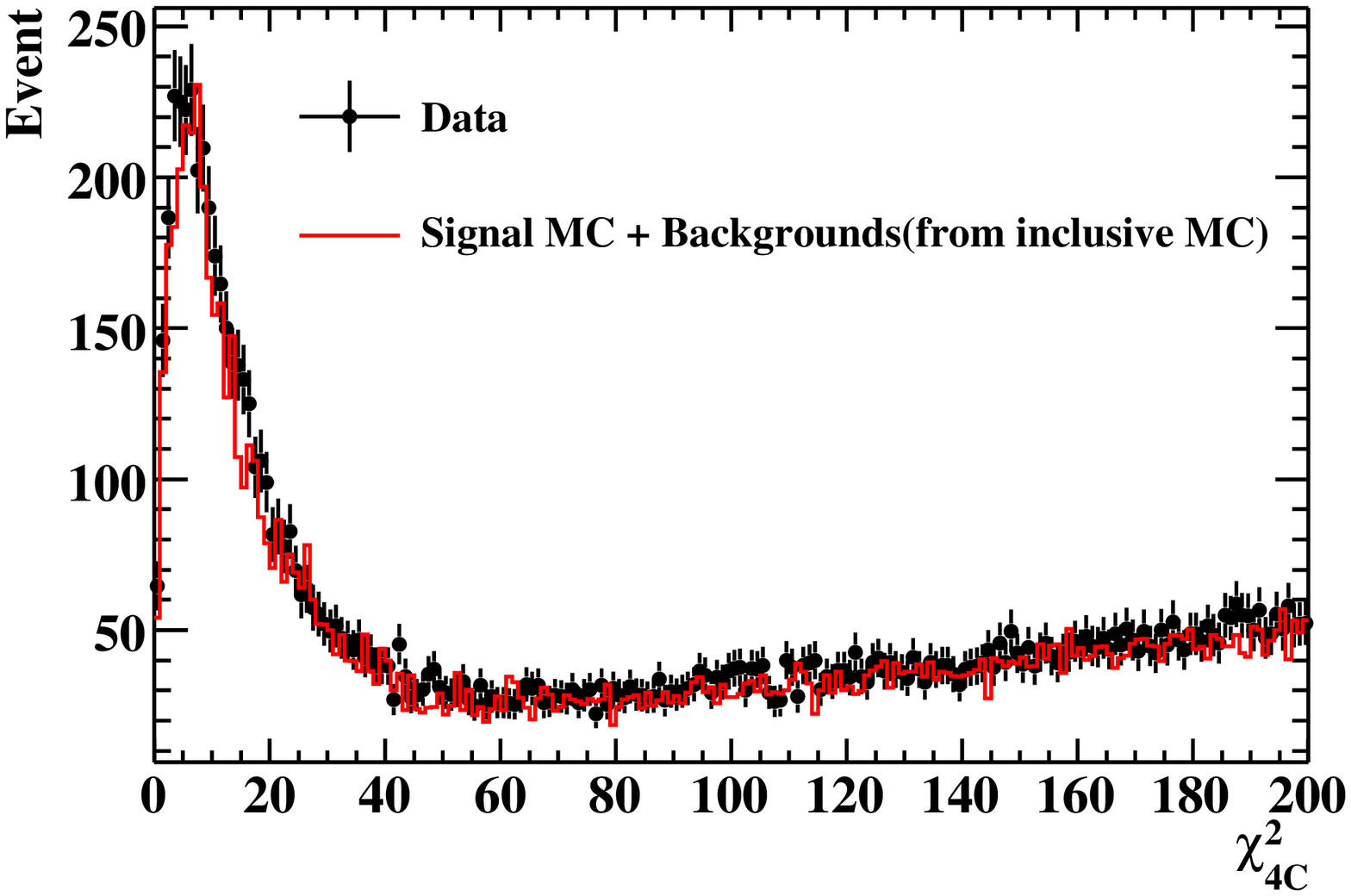}
\caption{Comparison of $\chi^{2}_{\rm 4C}$ between signal MC and
data for $\psp\to \gamma\ccj$, $\ccj\to \kkppp$. The points with
error bars are data, and the solid lines are signal MC plus
possible background estimated using inclusive MC sample. Left
panel: signal MC without track-parameter-correction; Right panel:
signal MC after track-parameter-correction. From top to bottom are
$\ccz$, $\cco$, and $\cct$, respectively.} \label{fig:sys compare
chisq kkpi}
\end{figure*}

The systematic errors in $\BR(\ccj\to \kskppp)$ ($\BR(\ccj\to
\kkppp)$) are 1.5\%, 1.9\%, and 1.7\% (0.4\%, 0.4\%, and 0.2\%)
for $J=0$, 1, and 2, respectively; and 1.5\%, 1.7\%, and 1.8\%
(0.9\%, 0.2\%, and 0.4\%) systematic uncertainties are assigned to
$\BR(\ccj\to \etac\pp)$ with $\etac\to \kskp$ ($\kkp$) for $J=0$,
1, and 2, respectively.

\subsection{Uncertainty from damping factor}

In the fit to the invariant mass spectrum of $\kskppp$ and
$\kkppp$, the damping function used by KEDR is adopted.
Another damping factor used by CLEO~\cite{CLEOdamp} is
$e^{-E_{\gamma}^{2}/8\beta^{2}}$ with $\beta=(0.0650\pm
0.0025)~\gev$ determined from their fit. Using this damping function
with $\beta=(0.097\pm 0.024)~\gev$ which is extracted from fitting
$\ccz$ data, the differences on the branching fractions of
$\ccj\to \kskppp$ ($\ccj\to \kkppp$) are assigned to the
systematic error due to damping function, which are 0.5\%, 0.1\%,
and 0.1\% (0.4\%, 0.1\%, and 0.1\%) for $J=0$, 1, and 2,
respectively. The effect for $\cco$ and $\cct$ is small since the
two states are very narrow.

\subsection{Uncertainty from intermediate states}

The detection efficiencies for the measurement of $\BR(\ccj\to
\kskppp)$ and $\BR(\ccj\to \kkppp)$ are estimated using the MC
simulation with $\ccj$ decay to $\kskppp$ and $\kkppp$ generated
according to pure phase space distribution. From the data, we
see broad intermediate states such as $K^{*}$ and $\rho$ in the
invariant mass spectra of $K\pi$ and $\pi\pi$. The branching
fractions of $\ccj\to \kskppp$ ($\ccj\to \kkppp$) via these
intermediate states are measured by fitting the invariant mass
spectra of $K\pi$ and $\pi\pi$. An alternative signal MC sample is
generated with all possible intermediate states and corresponding
branching fractions to determine the efficiency. The efficiency
difference between this sample and the phase space sample is about
1.0\% for $\ccj\to \kskppp$ and 4.0\% for $\ccj\to\kkppp$; these
are taken as the systematic error due to intermediate states.

\subsection{Uncertainty from fitting}

The systematic uncertainty due to the fitting range is estimated by
fitting the invariant mass spectrum in the range $3.25~\gevcc\sim
3.61~\gevcc$ and $3.35~\gevcc\sim 3.60~\gevcc$. The biggest
differences in the branching fractions are assigned as errors, which
are 1.0\%, 0.4\%, and 0.2\% for $\ccz$, $\cco$, and $\cct$,
respectively, in $\kskppp$ decay; and 0.4\%, 0.4\%, and 0.7\% for
$\ccz$, $\cco$, and $\cct$, respectively, in $\kkppp$ decay. The
background shape is changed from a second-order Chebyshev polynomial
function to a third-order Chebyshev polynomial function, and the
differences are taken to be the systematic errors, which are 1.4\%,
0.7\%, and 0.6\% for $\ccz$, $\cco$, and $\cct$, respectively, in
$\kskppp$; and 1.3\%, 0.7\%, and 0.4\% for $\ccz$, $\cco$, and $\cct$
in $\kkppp$.

\section{Results and discussion}

Using the numbers of signal $\ccj$ events from the fits, together
with the corresponding efficiencies, the branching fractions of
$\ccj\to \kskppp$ ($\ccj\to \kkppp$) are determined and listed in
Table~\ref{table:final results kk3p}. In the branching fractions of
$\ccj\to\kkppp$, contributions from narrow resonances
$\eta$, $\omega$, and $\phi$ are subtracted.  All these are first
measurements, and the branching fractions are at the 1\% level.
Comparing the two decay modes, we found the ratio of the branching
fractions is around one-half which may be a consequence of isospin
symmetry. We also measured the branching fractions of $\ccj\to\omega\kk$
and $\ccj\to\phi\pp\piz$ for the first time, the results are
listed in Table~\ref{table:final results kk3p} also.

\begin{table*}[htbp]
\caption{The results for $\BR(\ccj\to \kskppp)$, $\BR(\ccj\to
\kkppp)$, $\BR(\ccj\to\omega\kk)$, and $\BR(\ccj\to\phi\pp\piz)$.
The first errors are statistical and the second ones
are systematic.}\label{table:final results kk3p}
\begin{center}
\begin{tabular}{c|ccc}
\hline {Decay mode}     & $N^{\rm signal}$ & $\epsilon$ (\%) &
$\BR$ ($\times 10^{-3}$)
\\\hline
 $\ccz\to \kskppp$   & $2789\pm66$  & $9.30$          & $4.22\pm0.10\pm0.43$    \\
 $\ccz\to \kkppp$    & $9031\pm132$ & $10.34$         & $8.61\pm0.13\pm0.94$    \\
 $\ccz\to\omega\kk$  & $1414\pm42$  & $8.04$          & $1.94\pm0.06\pm0.20$     \\
 $\ccz\to\phi\pp\piz$& $538\pm29$   & $9.16$          & $1.18\pm0.07\pm0.13$      \\\hline
 $\cco\to \kskppp$   & $5180\pm75$  & $10.21$         & $7.52\pm0.11\pm0.79$    \\
 $\cco\to \kkppp$  & $12256\pm127$  & $11.10$         & $11.46\pm0.12\pm1.29$    \\
 $\cco\to\omega\kk$  & $628\pm29$   & $9.34$          & $0.78\pm0.04\pm0.08$     \\
 $\cco\to\phi\pp\piz$& $373\pm26$   & $10.50$         & $0.75\pm0.06\pm0.08$      \\\hline
 $\cct\to \kskppp$ & $4559\pm71$    & $9.76$          & $7.30\pm0.11\pm0.75$    \\
 $\cct\to \kkppp$  & $11189\pm124$  & $10.48$         & $11.69\pm0.13\pm1.31$    \\
 $\cct\to\omega\kk$  & $512\pm27$   & $8.58$          & $0.73\pm0.04\pm0.08$     \\
 $\cct\to\phi\pp\piz$& $408\pm28$   & $9.88$          & $0.93\pm0.06\pm0.10$      \\\hline
\end{tabular}
\end{center}
\end{table*}

With the upper limit on the numbers of events at the $90$\% C.L.  in
$\ccj\to \etac\pp$, $\etac\to \kskp$ ($\etac\to \kkp$), as well as the
corresponding efficiencies, the upper limits on the branching fraction
of $\ccj\to \etac\pp$ in the two decay modes are determined, as listed
in Table~\ref{table:final results kkp}.  We give a more stringent
constraint on the $\cct\to\etac\pp$ branching fraction than BaBar
does~\cite{chic2Babar}. The theoretical prediction of $\BR(\cco\to
\etac\pp)$ is also listed in Table~\ref{table:final results kkp},
which is larger than our measurements. We note that the theoretical
prediction uses experimental results as input to normalize the
parameters in the model. For example, the parameter
$\alpha_{M}/\alpha_{E}$ is extracted by comparing the branching
fraction of $\psp\to\hc\piz$ between the theoretical calculation and
the experimental measurement. This makes the prediction highly
dependent on the former experimental results and theoretical models.

\begin{table*}[htbp]
\caption{Upper limits at the $90$\% C.L. on $\BR(\ccj\to
\etac\pp)$ in the two $\etac$ decay modes. $N^{\rm fit}$ is the
number of events from the fits shown in Fig.\ref{fig: invariant mass kkpi}.
In $\cct$ case, $N^{\rm fit}$ includes the contribution from the peaking
background $\psp\to\pp\jpsi, \jpsi\to\gamma\etac, \etac\to\kskp(\kkp
)$.  }\label{table:final
results kkp}
\begin{center}
\begin{tabular}{c|cccc|c}
\hline {Decay mode}     & $N^{\rm fit}$    & $N^{\rm up}$ & $\epsilon$ (\%) &
$\BR^{\rm up}(\ccj\to\etac\pp)$ (\%) & $\BR^{\rm
theory}(\ccj\to\etac\pp)$ (\%) \\\hline
$\ccz\to (\kskp)\pp$ & $0.0\pm4.6$ & $6.8$    & $6.29$          & $0.07$    & - \\
$\ccz\to (\kkp)\pp$  & $0\pm15$   & $33.6$   & $6.82$          & $0.41$    & -
\\\hline
$\cco\to (\kskp)\pp$ & $18\pm17$  & $48.7$   & $9.45$          & $0.32$    & \multirow{2}{*}{$1.81\pm0.26$} \\
$\cco\to (\kkp)\pp$  & $6\pm25$ & $50.0$   & $9.82$          & $0.44$    &
\\\hline
$\cct\to (\kskp)\pp$ & $77\pm19$ & $64.1$   & $7.72$          & $0.54$    & -   \\
$\cct\to (\kkp)\pp$  & $89\pm26$ & $105.4$  & $7.83$          & $1.23$    & -
\\\hline
\end{tabular}
\end{center}
\end{table*}

\acknowledgments
The BESIII collaboration thanks the staff of BEPCII and the computing center for their hard efforts.
This work is supported in part by the Ministry of Science and Technology of
China under Contract No. 2009CB825200; National Natural Science Foundation
of China (NSFC) under Contracts Nos. 10625524, 10821063, 10825524,
10835001, 10935007, 11125525; Joint Funds of the National Natural Science
Foundation of China under Contracts Nos. 11079008, 11179007; the Chinese Academy
of Sciences (CAS) Large-Scale Scientific Facility Program; CAS under Contracts
Nos. KJCX2-YW-N29, KJCX2-YW-N45; 100 Talents Program of CAS; Istituto Nazionale
di Fisica Nucleare, Italy; Ministry of Development of Turkey under Contract No.
DPT2006K-120470; U. S. Department of Energy under Contracts Nos. DE-FG02-04ER41291,
DE-FG02-91ER40682, DE-FG02-94ER40823; U.S. National Science Foundation; University
of Groningen (RuG); the Helmholtzzentrum fuer Schwerionenforschung GmbH (GSI),
Darmstadt; and WCU Program of National Research Foundation of Korea under Contract No. R32-2008-000-10155-0.

\end{document}